**The role of austenite twins on variant selection during bainitic decomposition of a low carbon (0.06 wt%) steel**

Ruth Birch[1*], Ben Britton[1], Warren J Poole[1]

1. Department of Materials Engineering, University of British Columbia, Frank Forward Building, 309-6350 Stores Road, Vancouver, BC, Canada V6T 1Z4

*corresponding author: ruth.birch@ubc.ca

## Abstract

Thermomechanical Controlled Processing (TMCP) is widely used to control the microstructure and properties of line pipe or high strength low alloy steels (HSLA). These steels are often joined by welding and used in demanding environments such as the Arctic. In these materials, the thermal path the steel experiences is critical for understanding microstructural evolution during processing. A key step is the solid-state phase transformation during cooling from the high-temperature austenite to the room-temperature microstructure which significantly influences the final mechanical properties. We used 3D electron backscatter diffraction (EBSD) to explore the relationship between the austenite phase and the room temperature microstructure. A significant result for the present work is the collection, and analysis, of data from a large volume (150 x 150 x 100 $\mu m^3$, with a $(200\ nm)^3$ voxel size) which enables analysis of a complete prior austenite grain. This grain is twinned, allowing us to analyse the variants at the twin boundary in this grain, which offers new insights into the mechanisms of the transformation to the low temperature phase by highlighting the significant of the twin boundaries on the variants present. This suggests opportunities to engineer novel microstructures by controlling the high-temperature grain boundary character.

**Graphical abstract**

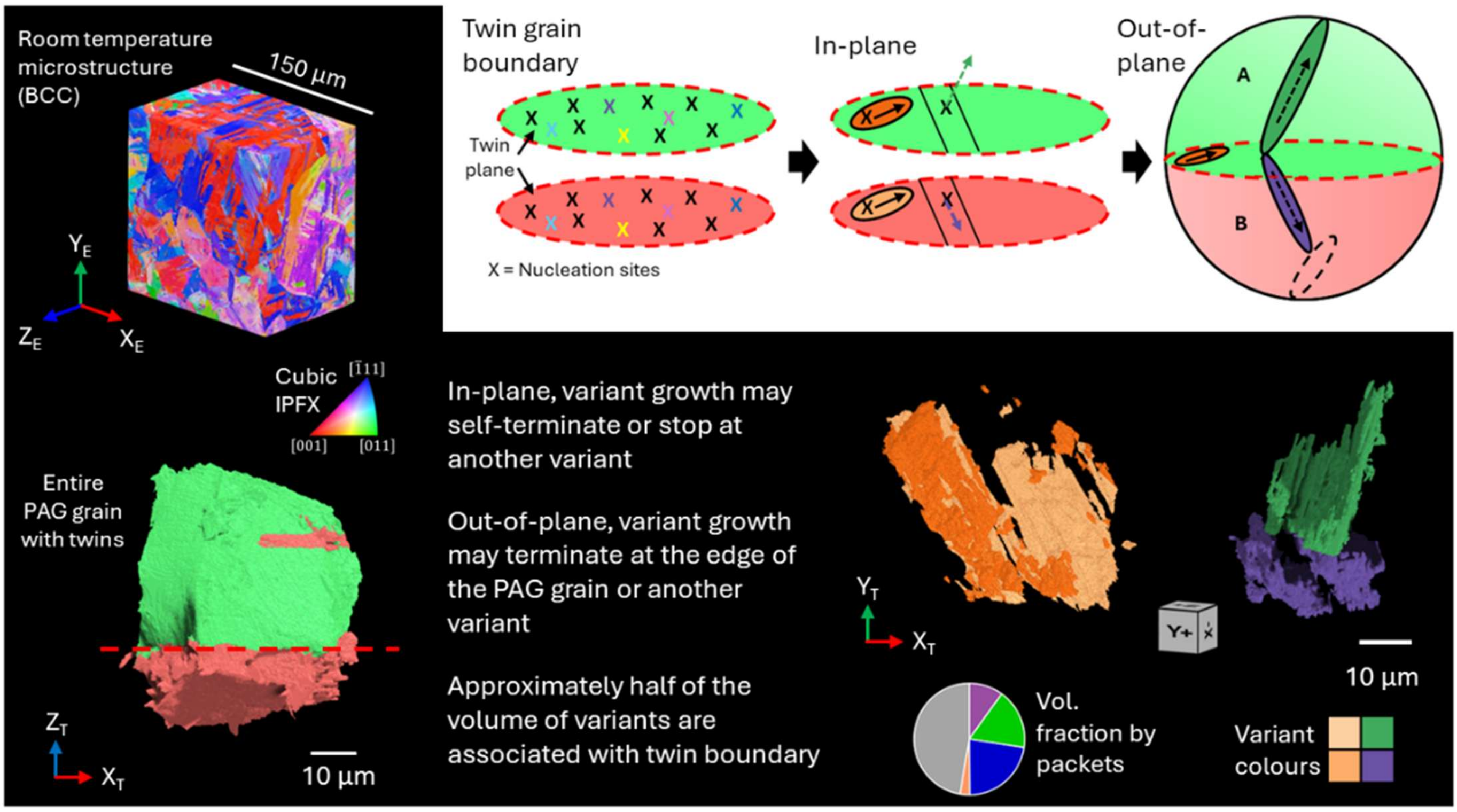

## 1 Introduction

Thermomechanical Controlled Processing (TMCP) is widely used to control the microstructure of line pipe and high strength low alloy (HSLA) steels. These steels are used for welded pipelines in demanding environments such as the Arctic, where high strength, weldability, and low temperature toughness are essential. A key step impacting the final microstructure is the solid-state phase transformation during cooling from the high-temperature austenite phase to the room-temperature microstructure, particularly at high cooling rates associated with welding or accelerated cooling after hot rolling. Here, complex bainitic or martensitic microstructures [1], are produced which significantly influence the final mechanical properties. Specifically, the population and morphology of the different child variants that form affect the grain boundary network in the final microstructure, which influences the strength and toughness of the final component [2]. Certain boundary types in the high temperature austenite microstructure (e.g. high angle, low angle or special boundaries such as annealing twins) may have a specific influence on variant selection in the final microstructure [3]. The potential therefore exists to tailor the mechanical properties of the steel to the application if one has an improved understanding of the microstructural evolution at each step. With the availability of three-dimensional electron backscatter diffraction (3D EBSD) experiments, we can now look at variant selection within a complete prior austenite grain.

The high temperature, austenite microstructure, whilst difficult to observe directly, can be reconstructed from the final room temperature bainitic or martensitic microstructure using the orientation relationship (OR) between the two phases. Common ORs used for steel are Kurdjumov-Sachs (KS) where $(111)_\gamma//(011)_\alpha$ and $[\bar{1}01]_\gamma//[\bar{1}\bar{1}1]_\alpha$, and Nishiyama-Wasserman (NW) where $(111)_\gamma//(011)_\alpha$ and $[1\bar{1}0]_\gamma//[100]_\alpha$) [4]. Whilst these two ORs are similar, there is a rotation about the closest packed plane, {111}, of 5.26° [5] and as a result the main difference is that KS gives 24 unique child variants compared with 12 for NW. For any material with a solid-state phase transformation, it is these variants, and the preferential selection of them that can impact the ultimate mechanical properties. For example, there may be coarse microstructure if there is a limited number of variants within a prior austenite grain; and in contrast if a large number of different variants are generated, there can be a finer final microstructure with a greater number of higher angle grain boundaries. Evidence from the literature indicates that the finer microstructures with a high number of different variants can strength, ductility and resistance to crack propagation [3].

Studies of variant selection in steels have included evaluating factors such as the effect of stress [6] or deformation at different points during processing [7] and the effect of prior austenite grain size [8]. Variant selection at prior austenite grain boundaries has been considered in terms of nucleation rules for these boundaries [9], and it has also been suggested that special grain boundary types, such as twins, can play a role.

Annealing twins are a common occurrence in face centred cubic (FCC) materials with low to medium stacking fault energies [10], including low-carbon steels in the austenite phase field. These annealing twins usually take the form of a 60° rotation about <111> (i.e. a Σ3 twin) [11] and have been observed to form during austenite grain growth [12]. For the variants, this means that there is a common packet group (variants with the same $[111]$ type direction in the parent grain) between the parent grains, which may influence variant selection at the twin boundary. *In situ* experiments using laser scanning confocal microscopy to look at bainite formation have concluded that the presence of twins impacts bainite growth and morphology [12], and that the twin boundaries can stop bainite growth and provide potential nucleation sites [13].

The presence of twins complicates the reconstruction of the parent phase using EBSD data and reconstruction codes. This is because, when the OR is strictly obeyed, the child orientations (variants) in the shared packet can be associated with either parent grain, making it very difficult to group the data into the correct parent grain. This problem is particularly significant if these shared variants are located at the boundary between the grains. However, experimentally, a slight rotation away from the strict ORs can be observed, and this helps differentiate between variants associated with each prior austenite grain (PAG). For more discussion on twins in reconstruction and methods to address this challenge, see [14].

Whilst most studies are based on 2D metallographic or EBSD observations, with the development of 3D systems, for example, using focused ion beam – scanning electron microscopes (FIB-SEMs) [15]–[20], research on steels has moved into 3D, opening up opportunities to study grain size and morphology in greater depth.

For example, Petrov et al. [21], [22] have used small targeted volumes (e.g. max. 16x14x6 μm$^3$ at 0.1 μm$^3$ voxel size) to investigate grains and texture in pipeline steels. Parida et al. [23] looked at texture and recrystallisation in modified 9Cr–1Mo steel using two volumes, adapted to the grain size before and after recrystallisation (e.g. area per slice 4x4 μm$^2$ vs. 30x10 μm$^2$). Mosayebi et al. [24], [25] evaluated a triplet of (partial) PAGs in low carbon martensitic steel, covering a volume of 135x120x75 μm$^3$ at 0.5 μm$^3$ voxel size for a sample with average PAG size of 158.1 μm. Their analysis includes evaluation of variants and packet groups within the grain, and they observed the 3D arrangement of the packets in a tetrahedral arrangement within the PAGs and calculated a dominant habit plane of $\{557\}_\gamma$. Finally, Shibata et al. [26] looked at medium carbon steel with a prior austenite grain size of 122 μm and fine lath microstructure. They imaged a small volume in 3D, measuring 20x20x6 μm$^3$, with a voxel size of 6.25x6.25x5 nm$^3$, and investigated the local variant configuration by combining the 3D dataset with 2D EBSD and TEM analysis.

In this work, we carried out a 3D EBSD tomography experiment on a line pipe steel to collect a 150 x 150 x100 μm$^3$ volume with a voxel size of (200 nm)$^3$ containing a fully internal, twinned prior austenite grain, with the aim of evaluating the variants at this twin boundary and the impact of the twin boundary on variant distribution within the grain.

## 2 Experimental

### 2.1 Sample

A line pipe steel with the composition at Table 1 was used for this tomography experiment. Using a DSI 3500 thermomechanical simulator, the 100 x 11 x 11 $mm^3$ sample was heated to 1150 °C at 50 °C/s with a hold at 1150 °C for 10 mins. During the heat treatment, laser ultrasonic measurements (see [27], [28], with the calibrations based on [29]) were used to measure the prior austenite grain size of ~40 μm. Then the sample was cooled to room temperature at a nominal cooling rate of 50 °C/s until austenite decomposition was initiated. This heat treatment resulted in a bainitic microstructure, as discussed in 2.2. To perform tomography, the sample was sectioned using wire electrical discharge machining (EDM) to approx. 3 x 3 x 5 $mm^3$ to enable the tomography experiment - see supplementary material for more details on geometry.

*Table 1 – Composition of line pipe steel used in this work (in wt%).*

| C | Mn | Cu+Ni+Cr+Mo | V+Nb+Ti |
|---|---|---|---|
| 0.06 | 1.3 | <0.9 | 0.07-0.08 |

## 2.2 Sample microstructure

It is noted that in low carbon steels (0.06wt% carbon in the current case), there is continuous spectrum of decomposition products evolving from polygonal ferrite to acicular ferrite (or Widmanstätten ferrite) to bainite to martensite as the cooling rate increases (or equivalently as the undercooling is increased) with little if any iron carbides observed in bainite or martensite. In the current study, the transformation start temperature was experimentally measured to be ~575°C and the transformation finish temperature was 460°C. It is noted that the cooling rate reduces to ~12.5 °C/s at 550 °C due to the latent heat of the transformation and then rebounds to ~30 °C/s at 450 °C. The equilibrium $T_{ae1}$ and $T_{ae3}$ temperatures were calculated to be 670 °C and 851 °C using Thermocalc (TCFE7 database) and the bainite start temperature was estimated to 645 °C, see reference [30]. Further, the martensite start temperature was estimated to be in the range of 465-470 °C from the equations of Andrews and Steven-Hayes [31], [32]. As such, it was summarized that austenite decomposition led to a bainitic final microstructure with a fine lath structure which will be shown later (additional surface EBSD maps are also provided in supplementary figure S2).

## 2.3 Tomography

A 3D EBSD serial sectioning ('tomography') experiment was carried out using a TESCAN AMBER X plasma focused ion beam – scanning electron microscope (pFIB-SEM) equipped with an Oxford Instruments symmetry S2 EBSD detector and a TESCAN rocking stage. This instrument has a static EBSD setup, where the sample is facing the EBSD detector at a 70° tilt, but rotated so that the top of the edge to cut is perpendicular to the FIB beam (see Figure 1). The static setup is described in more detail in [19] - for this experiment, tomography was performed on the sample edge (i.e. no lift out step).

A schematic for the target volume preparation is shown in Figure 1 (A). A 225 x 200 x 15 $\mu m^3$ (W x L x D) tungsten cap was deposited using a (FIB) voltage of 15 kV and current of 120 nA. To allow the required exit angles and minimise shadowing, two triangular side trenches are cut, as shown in the schematic. These were milled using (FIB) voltage of 30 kV and current of 1 μA – in this case, the chosen region of interest (ROI) was located at the edge of the sample, so one of the trenches is rectangular, as shown in the Figure 1 (A). The sides and front face were then 'cleaned up' (polished) using a 30 kV 300 nA polish ready for tomography.

Each slice was cut using a (FIB) voltage of 30 kV and current of 300 nA with a FIB slice thickness of 0.2 μm. The rocking stage y-tilt was used to tilt the sample +6 °/-1 ° during the milling (Figure 1 (C)), with 2 rocking cycles used per cut, again to reduce curtaining artifacts during milling. Imaging and analytics were always collected at y-tilt 0 °. Unique to this microscope and experimental geometry is the fact that the sample was aligned using the electron and ion beams during the FIB tomography 'wizard'; and the EBSD detector and sample are within the so-called 'static' configuration. Together, these both significantly reduces misalignment during the 3D experiment, which was validated using preliminary experiments. To further improve alignment, two alignment markers were used: (1) for the FIB milling, an alignment marker was milled on the top of the bulk material behind the target volume; (2) for SE- and BE-imaging, and EBSD, a second alignment marker was milled by the side trench. Verification of the alignment was performed via analysis of the SE- and BE- imaging sequences, and a preliminary sectioning experiment. Slice-to-slice variations are expected to be sub-voxel (i.e. less than 200 nm). For the purpose of the analysis here, no further alignment was deemed necessary.

During the tomography, an EBSD map was captured for each slice using a voltage of 20 kV and a current of 10 nA for an area of 150 x 150 $\mu m^2$ at a step size of 0.2 μm. Patterns were captured at 156 x 128 pixels with an exposure time of 0.5 ms and processed online in Oxford Instruments AZtec software, with a minimum of 6 bands required to index the point.

Each slice took ~12 mins to complete (FIB slicing 5 m 42s, SE Imaging 19 s, EBSD ~5 mins, +overhead). Whilst 700 slices were collected, to address issues at the start and the end of the run, only a subset of 500 slices were used for this analysis. The final analysis volume was 150 x 150 x 100 $\mu m^3$ at a voxel size of $(0.2\ \mu m)^{3,}$ as shown in Figure 1 (D).

The FIB and EBSD conditions used in this data collection were optimized via initial trial experiments. Specifically, the FIB-polishing steps used during sectioning were demonstrated to not cause significant problems for the reconstruction of the parent structure (i.e. no additional transformations were caused for the ferrite phase) and the EBSD step size resulted in enough data to successfully reconstruct the parent structure.

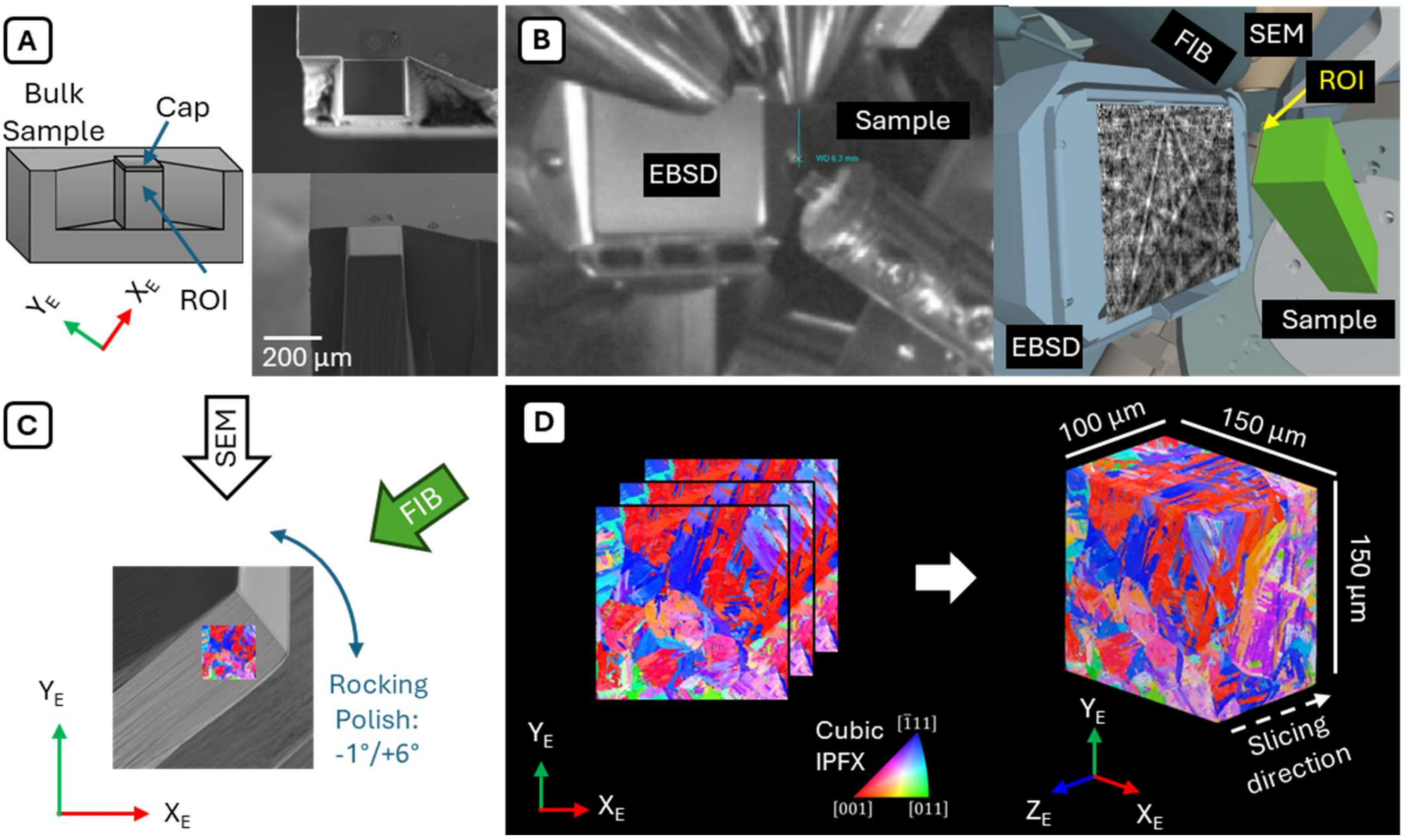

*Figure 1 – Tomography overview. (A) Sample Schematic and secondary electron (SE) images at the start of the experiment (Top = FIB – SE; Bottom = SE) Scalebar = 200 µm; (B) Chamber view and model view of 3D setup; (C) Details of the EBSD map area vs. sample geometry; (D) slices are combined to give the 3D volume – IPFZ.*

### 2.4 Analysis

Data analysis was carried out in MATLAB (2023b & 2024a) using the MTEX toolbox (5.11.2) [33], with visualisation and further analysis in Dragonfly (2024.1 & 2025.1) [34]. Code was developed to import MTEX-generated EBSD data for visualisation in Dragonfly, and to extract objects from Dragonfly for crystallographic analysis in MTEX (and vice versa).

Prior austenite grain reconstruction was carried out using in house code (3D PAG) which builds on prior work by Nyyssönen et al [35] and the authors of this paper [14]. In brief, the 3D PAG code reconstructed each slice in 2D, including 're-sort' [14] to improve the reconstruction. Subsequently, data in the volume was linked together (using every 10$^{th}$ slice as a seed for naming the parent grains). Then groups of 3 slices were considered at a time to provide a more consistent reconstruction throughout the volume, using a forward and backwards pass approach. Local analysis of the grain considered in this paper was carried out by restricting the problem to the grain region and its immediate neighbours, with an adapted re-sort approach. Note that, to avoid inconsistencies slice to slice, a mean iterative OR was initially calculated (and used throughout) by reconstructing every 10$^{th}$ slice in the volume and averaging the iteratively determined ORs (see [35]), where in each case Kurdjumov-Sachs OR was used as a starting point. The final OR used for the volume, in rounded Bunge Euler angles, was $(330,9,344)_{FCC\rightarrow BCC}$. Variant numbering used follows the convention from Morito et al. [36], and while the variant colors within each prior austenite grain in 3D are consistent, variant numbers between different prior austenite grains are necessarily unrelated.

The main reconstruction was run using Compute Canada resources, with post processing carried out on a local workstation computer. For the dataset presented, the volume was reconstructed, then a PAG was selected, and a secondary check/re-sort was run on this local grain region to improve the slice-to-slice reconstruction. When looking at this PAG region, to analyse the variants, 'connected component' analysis in Dragonfly was used on each variant in turn, and groups with less than 50 voxels were removed and this data was subsequently analysed in MTEX.

## 3 Results

### 3.1 Isolated Grain

A twinned PAG pair was isolated from the 3D volume as shown in Figure 2. The location of the PAG within the volume is indicated by arrows in Figure 2 (A) for the initial microstructure and (B) the reconstructed PAG microstructure. The local region to the PAG has had the additional processing (a version of re-sort [14] in 3D) to address small issues with the reconstruction where the code can consider too many similar PAG options locally and struggle to assign a PAG orientation – it was not computationally cost effective to expand this to the whole volume at this time. The complete isolated grain is then shown in Figure 2 (C) for the initial microstructure and (D) for the reconstructed microstructure. In all cases, a box representing the full volume is shown to help show the spatial location of the grain. For a 3D video showing the volume and the grain location, see the supplementary material.

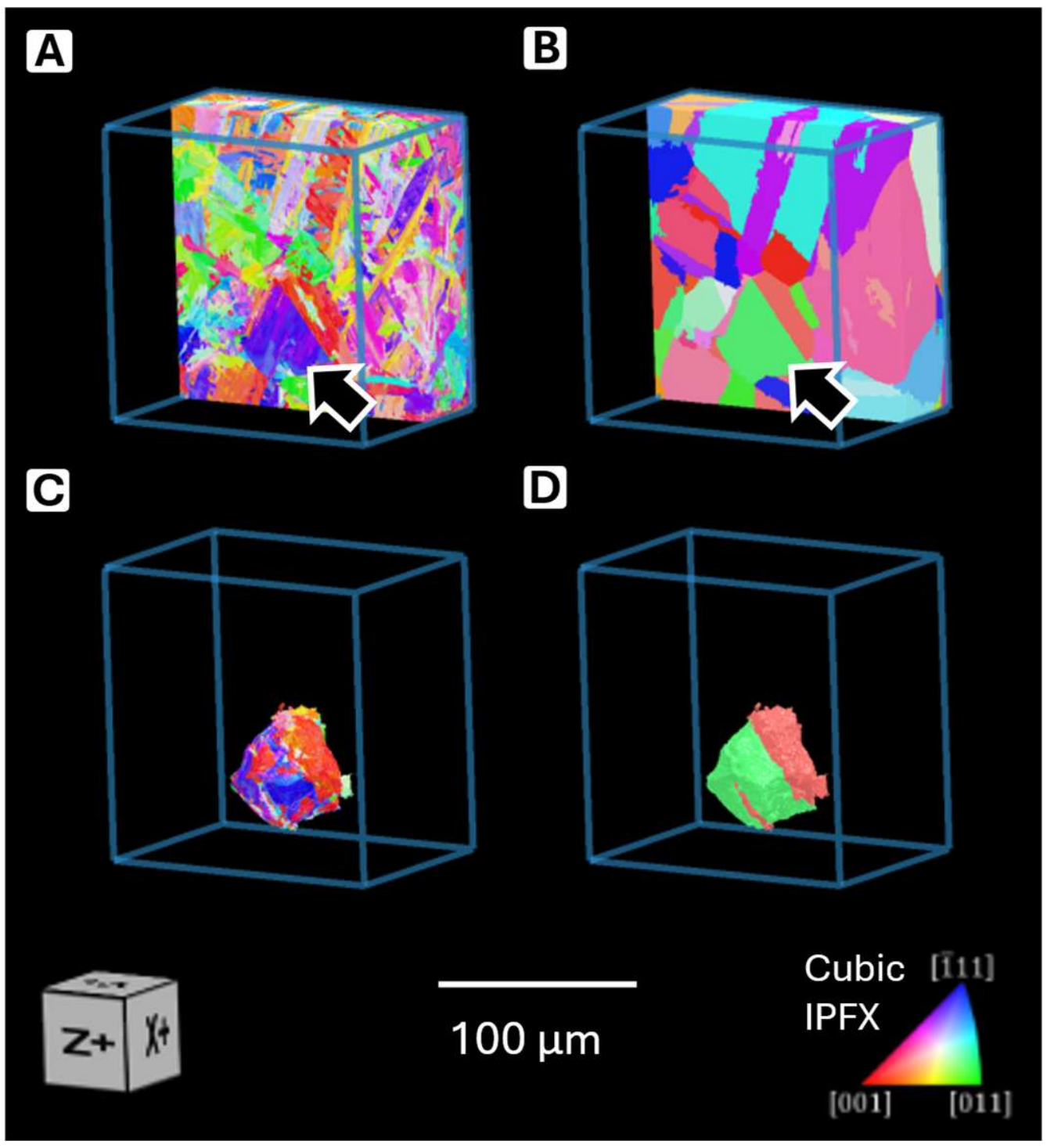

*Figure 2 – Cut-throughs showing the location of the isolated grain within the volume. (A) IPFX initial; (B) Reconstructed prior austenite grain (PAG) microstructure with the grain identified with arrows; (C) isolated grain – initial microstructure; (D) isolated grain – reconstructed PAG microstructure. The outline box indicates the full volume.*

The isolated grain pair contains a full and a partial twin, both parts with the same parent orientation. The analysis in this paper focusses only on the larger section of the twin and its boundary, with the smaller twin region excluded from the analysis. The larger grain (green in Figure 2 (D)) and the larger of the twin parts (orange in Figure 2 (D)) will be referred to as PAG A and PAG B, respectively, from this point.

The (rounded) mean grain orientations are (249°, 43°, 127°) for PAG A and (343°, 20°, 2°) for PAG B in the EBSD axis system, using Bunge Euler angle notation. The misorientation between the grains is 59.8° about $[\bar{1}11]$, using the mean orientation of the reconstructed parent grains. Additionally, a 5-parameter grain boundary analysis was carried out as shown in Figure 3 (A).

To calculate the grain boundary interface normal vector, the grain boundary trace was measured from approximately the centre of the grain pair and the perpendicular bisectors calculated. The grain boundary normal was calculated using the intersection of these two zones on the pole figure (Figure 3) and the normal vector was calculated as $[-0.72 \;\; -0.52 \;\; 0.34]$ in the sample frame.

Knowledge of this normal vector was used to rotate the crystallographic and spatial microstructure data so that the twin grain boundary plane could be viewed in the (viewing) XY plane. To do this, the data was rotated such that the normal was rotated along the XY zone to the centre of the pole figure and a second rotation was made to align the $[101]_{PAG\ B}$ with the X-axis. The results of this are shown in Figure 3 (B) – original and Figure 3 (C) – rotated with new axis system, along with cubes representing the orientations for the grains in each. For simplicity, the original IPFX colours are used for the grains throughout this paper.

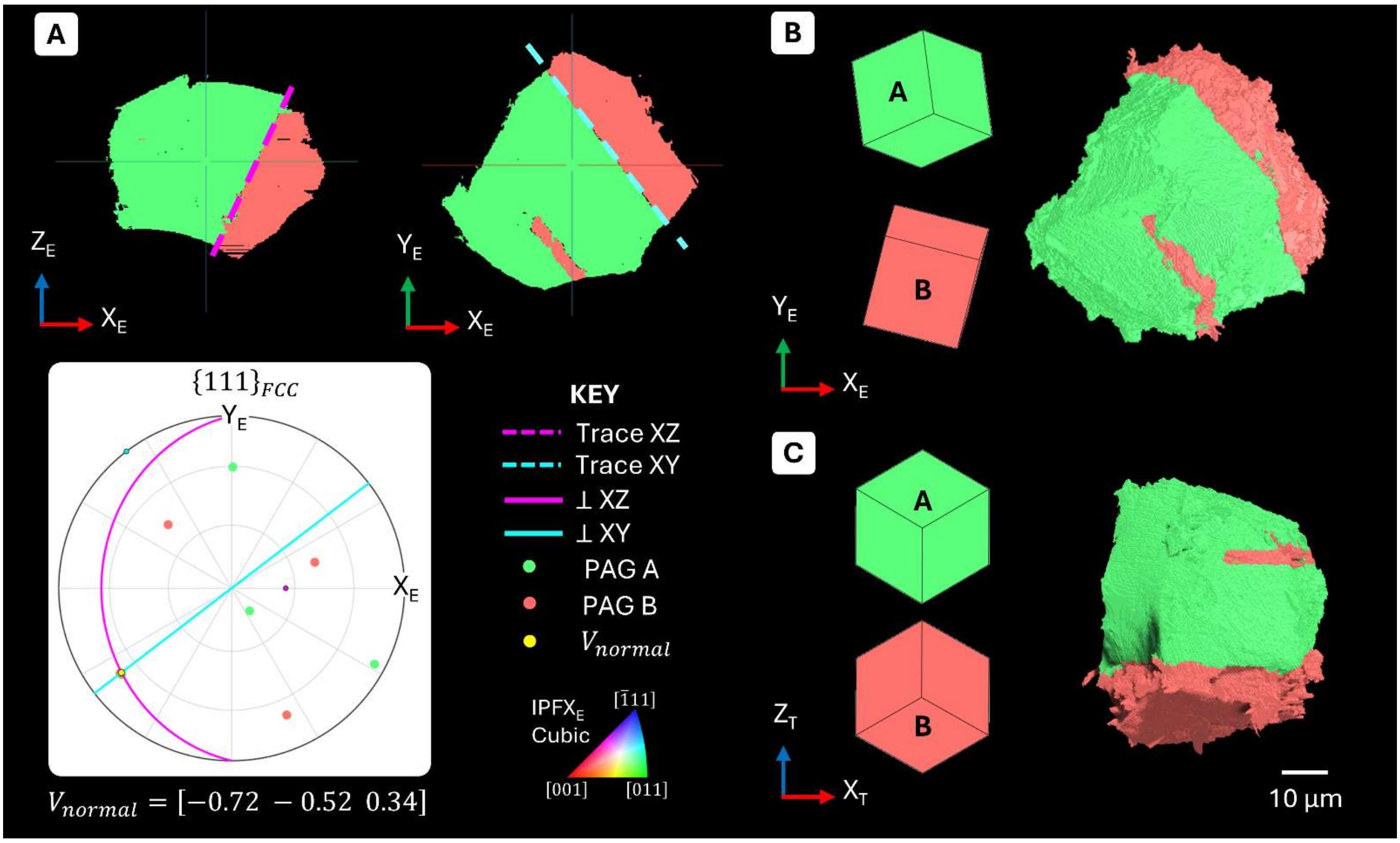

*Figure 3 - Isolated twinned grain (A) grain boundary analysis to identify the grain boundary normal – including XZ and XY planes through the centre of the grain with the trace of the grain boundary indicated by the dashed lines and pole figure analysis with the perpendicular bisectors overlaid and grain boundary normal identified; (B) Grain and orientation cubes in the EBSD orientation; (C) Grain and orientation cubes rotated so that the twin plane is the XY plane. IPFX (EBSD orientation) colours used throughout for the grains.*

### 3.2 Twin Boundary

To explore the twin boundary in more detail, a thin region centred on the twin boundary was isolated, as described in the schematic in Figure 4 (A). The region is 2 µm thick (10 voxels), with approximately the same number of voxels belonging to each PAG contained within. Figure 4 (B) shows the experimental results with (i) the complete grain pair, (ii) the thin region, separated by PAG association, and (iii) the twin plane at the intersection in both PAG IPF-X colour.

The reconstructed twin boundary is not flat. This could be found in practice, either due to stepping of the Σ3 interface that can consist of flat coherent interfaces and incoherent steps or due to sympathetic growth of variants in and across the parent-twin boundary. However, in practice there is the experimental complexity related to the data analysis and reconstruction routine, for example this could be due to some local difficulties with assigning points to each parent where the child orientations may belong to either parent grain. In practice, this roughness results in ‘gaps’ in the thin region when separated by PAG, as highlighted by the blue ellipses in Figure 4 (B-iii) and these are due to the voxels in that region being grouped with the other grain. For further discussion of the exact twin position and flatness of the interface, see supplemental information (and Figure S4), where we conclude that the position is sufficiently determined for our analysis.

The final part of this figure is the twin planes in variant colours (Figure 4 (B-iv)). These are the child orientations grouped into the closest of the 24 variants for each grain, as calculated using the mean OR for the volume. Two observations can immediately be made: (1) the variants in this region are in laths (sometimes referred to as 'sheaves'), at approximately 60° from each other, along the ⟨110⟩ directions in the PAG and (2) there are variants that appear in similar locations (spatially correlated) on both sides of the twin interface e.g. green in PAG A and purple in PAG B.

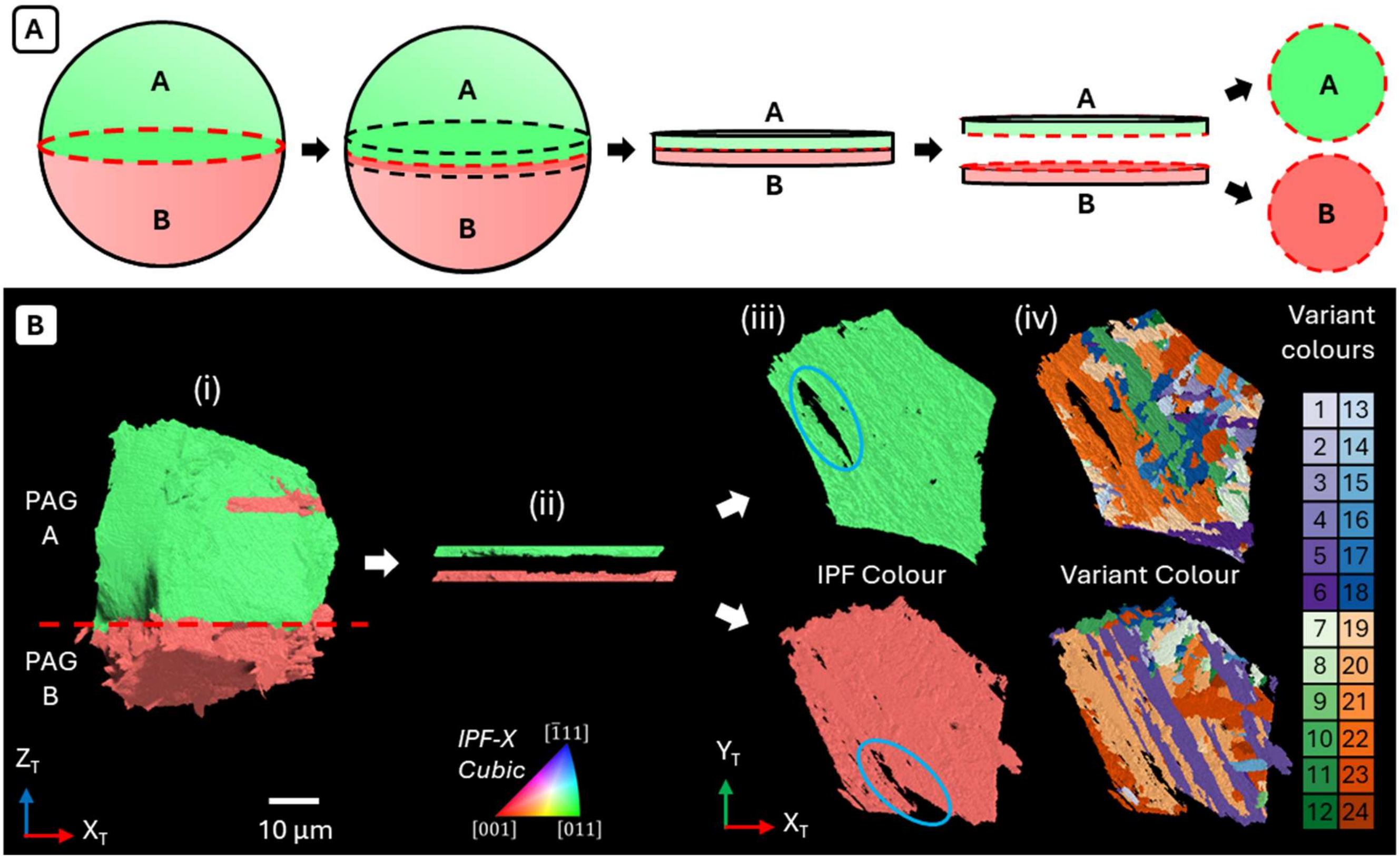

*Figure 4 – Twin plane analysis. (A) Schematic showing how a thin region around the twin grain boundary was analysed and the twin planes for each grain shown. (B) Experimental results in IPFX colours for the PAGs and the twin planes in variant colouring for each PAG. Note: For PAG A, the images at B(iii) and B(iv) are flipped in X compared to the schematic to make it easier to see the correspondence.*

### 3.3 Variant statistics

For statistical analysis of the variants, only variants with >50 voxels after connected components 3D (6-connected) analysis in Dragonfly are considered. All 24 variants are found in PAG A, whilst only 21 variants are found in PAG B. To evaluate the impact of variants found at the twin boundary on the grain as a whole, the distribution of variants was calculated for two regions, in each PAG: (1) the twin boundary region (Figure 4 (B)) and (2) the complete PAG.

An example is shown in Figure 5, using packet groups (where the variants have a common [111] direction in the parent grain) for the variants. Looking at the twin boundary only (Figure 5 (A)), we see that the shared packet group, indicated in orange in both PAG A and PAG B, is dominant in both grains, making up approximately 50% of this region in each case. The shared packet group is due to the twin being a ~60° rotation about a ⟨111⟩ direction, which leads to similar variant orientations in this PAG direction. The minimum misorientation angle between the individual variants in the shared packet is ~5°. This is large enough to help differentiate between the two potential PAG orientations in most cases, but may still lead to some uncertainty in grouping between PAGs – see supplemental material for further discussion. However, it is still clear that this is the dominant packet at the twin boundary for both grains.

Looking at PAG A, as we move into the grain, this proportion changes and other packet groups have higher distributions. In Figure 5 (B), only 13% of the voxels in the complete PAG are associated with the shared packet, indicating that these variants are mostly contained within the twin plane region. Note that this proportion may be impacted due to voxels within PAG A that are associated with the smaller partial twin PAG.

In PAG B, the situation is similar when considering the complete grain (Figure 5 (B)), we see a reduction in the proportion from 51% to 37%. However, in this smaller PAG, this packet group continues to dominate when considering the whole grain.

Looking at the other packet groups, for PAG A, the distribution starts ~equal for the packets, and then two packet groups become dominant (see Figure 5) indicating the growth of some preferred variants. In PAG B, this is more pronounced, with one dominant non-shared packet group (purple) in both cases. Both of these results show some preferential nucleation and growth of the variants with respect to packet groups.

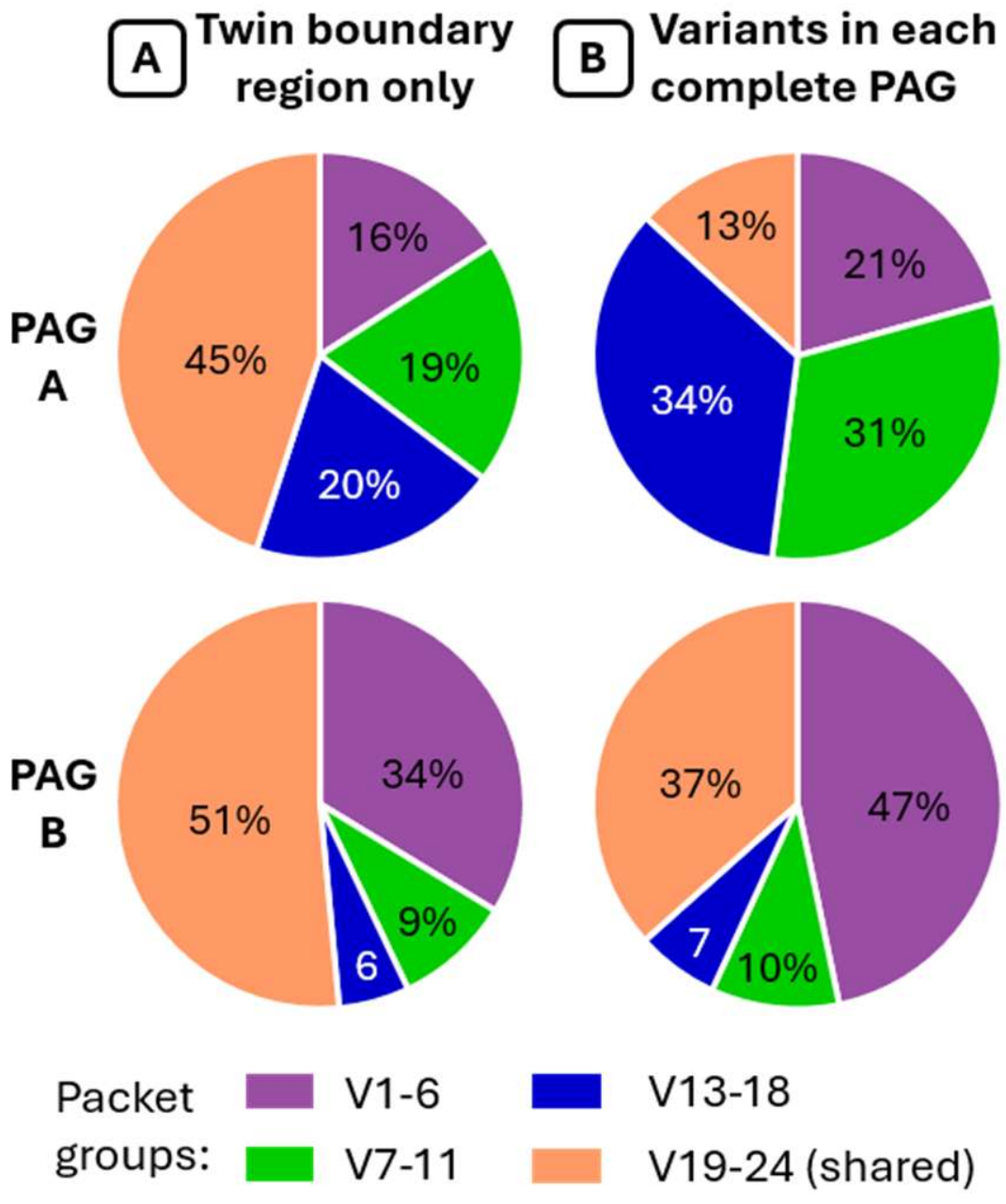

*Figure 5 - Variant distribution into packet groups. (A) twin boundary region only – 2 µm box around the twin boundary and (B) in the complete PAG volumes. Variants containing a minimum of 50 voxels included. Total voxels considered: PAG A: 4,984,565 (complete grain) / 176,047 (twin boundary region); PAG B: 1,868,047 (complete grain) / 163,087 (twin boundary region). Variant numbers are specific to the PAG – only the shared packet group variants are common (due to symmetry) and can be read across. Exact voxel counts are included in the supplemental material.*

Another way to consider the variants is to classify them into terms of Bain groups, where the $(001)$ plane of the variants are grouped around a $(001)$ plane in the parent orientation, resulting in three distinct groups of eight variants. When the same analysis as Figure 5 is done for Bain groups (see supplementary material), one Bain group dominates in each case for PAG B, with 80% of the twin boundary region voxels, 65% connected and 93% overall. In PAG A, there is also a dominance of one Bain group at the boundary (60%) but when considering the whole grain, a different Bain group is dominant at 53%, having been just 13% in the twin boundary region. Finally, the individual variant proportions can be evaluated, as will be discussed later (see supplementary material for figures).

### 3.4 Variants in 3D

The voxels associated with each variant were isolated and visualised using Dragonfly. For PAG A, the results are shown in Figure 6, where the variant colours are the same as in Figure 4. To aid understanding of this structure, a video rotating the 3D volumes together through 360° is found in the supplemental material. Here the value of 3D is clearly shown – we can start to see patterns in the morphology and crystallographic direction of the variant types. For example, in this view, we see that the bottom row (V19-24 / orange), which are variants in the shared packet, all have an alignment in a horizontal direction, which is in this case, the twin plane.

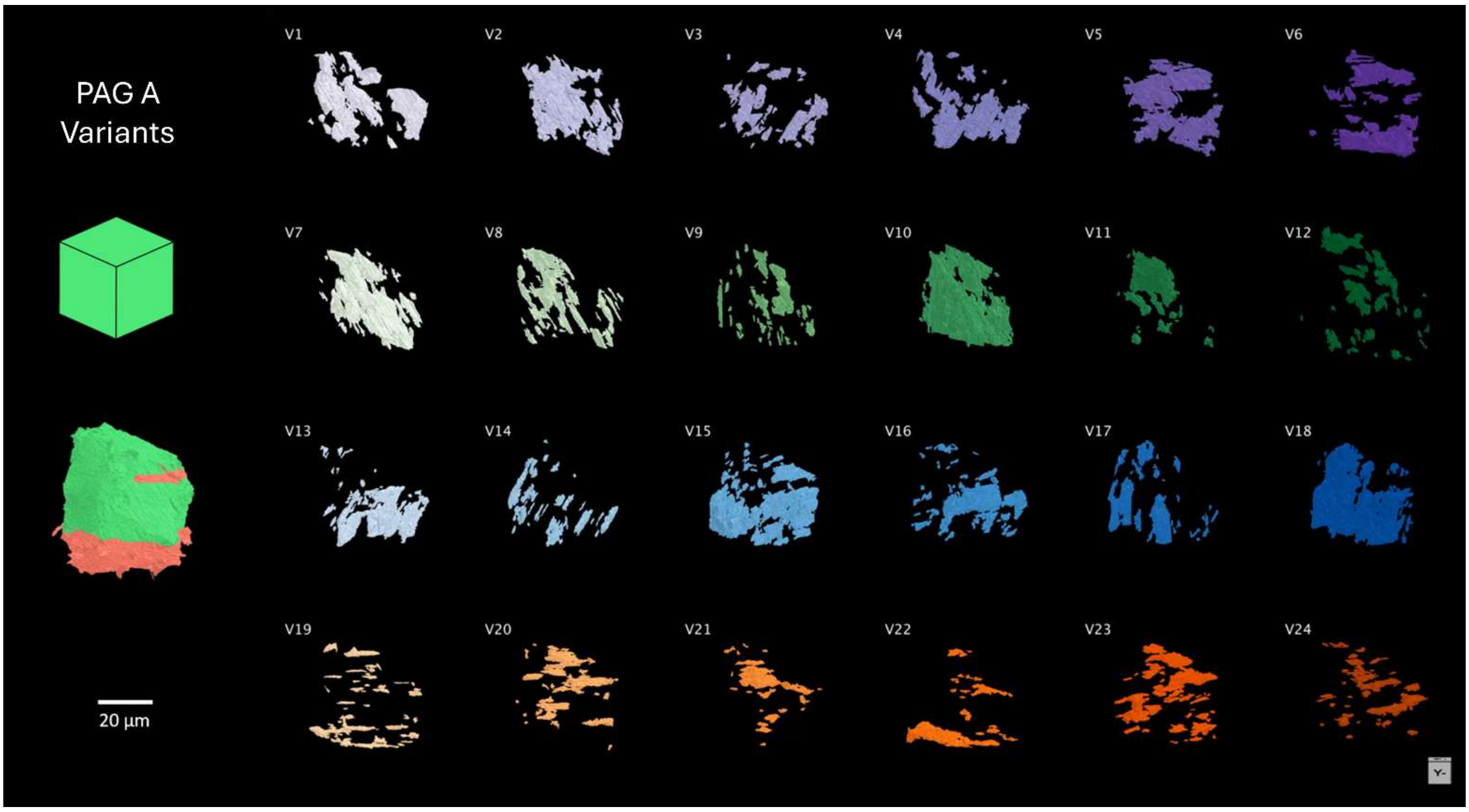

*Figure 6 - Variants for PAG A divided into the 24 variant types (regions less than 100 voxels have been removed). Variant colouring is the same as Figure 4. To aid understanding of this structure, the reader is directed to the supplemental material for a video where the 3D volumes rotate together.*

Isolated variant regions, defined as contiguous groups of voxels labelled as the same variant, were analysed in dragonfly by extracting their eigenvectors. An example is shown in Figure 7 (A) where a region (pink) is extracted from the all the voxels of the same variant for PAG A, and the principal axes shown. Due to the slightly irregular shapes of these regions, the directions for the long and short axes were analysed rather than the aspect ratio or sizes. Figure 7 (B) shows the distribution of directions for the eigenvectors of the longest and shortest principal axes for all PAG A isolated variant regions. Relevant crystallographic directions from PAG A were then overlaid (stars) to indicate the directions with which the principal axes are preferentially aligned. This analysis shows that the long directions of the variant regions tend to align with the $\langle 110 \rangle$ in the PAGs, whilst the shortest directions are typically aligned with the $\langle 111 \rangle$ in the PAGs.

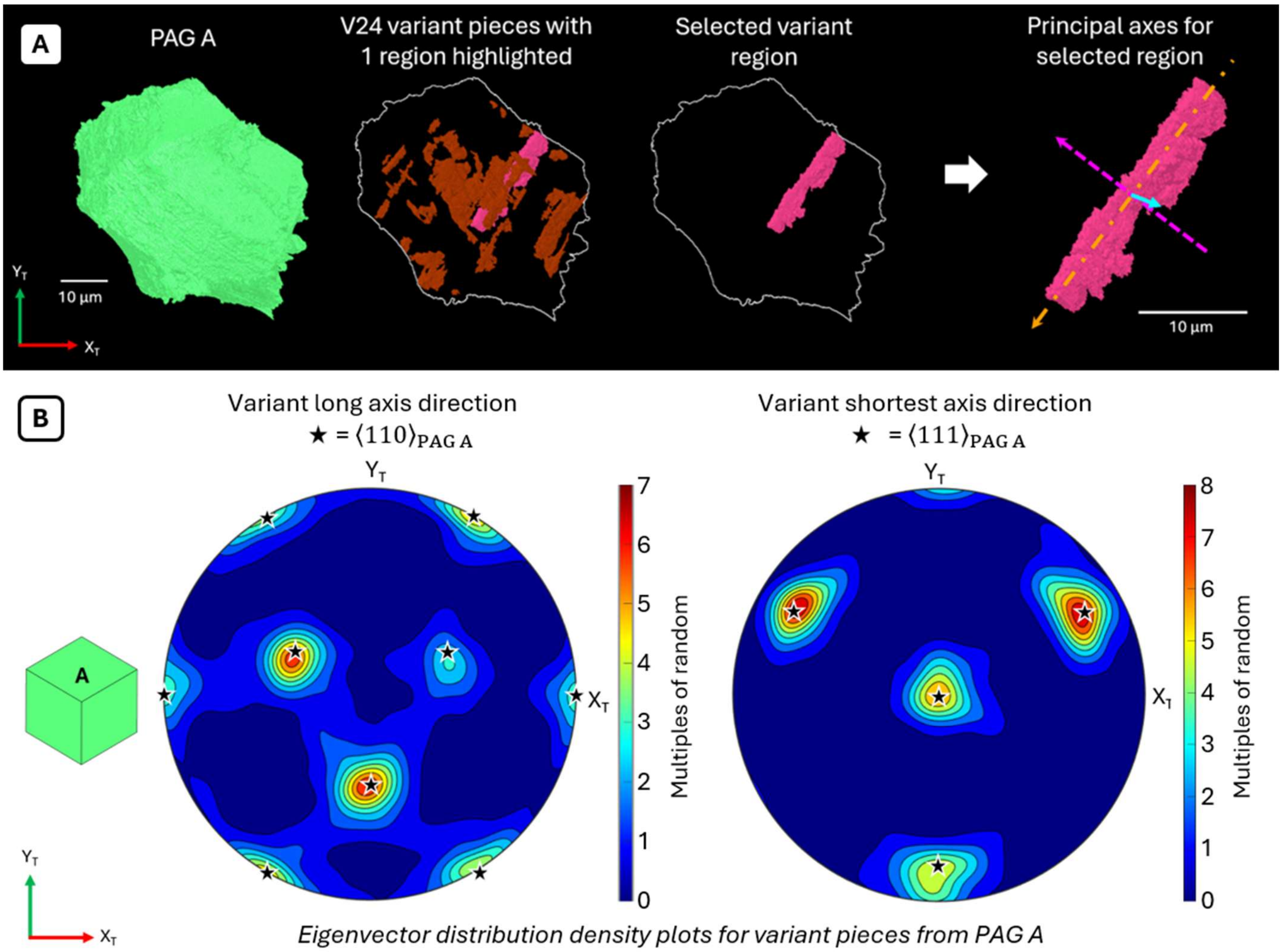

*Figure 7 - Variant direction analysis. (A) Example showing one isolated variant region for a PAG A variant and its principal axes. (B) Eigenvector distribution density plots created from the eigenvectors for PAG A isolated variant regions for their longest and shortest axes. Variant regions >100 voxels included. Halfwidth 6°.*

## 4 Discussion

In this paper, we have presented the results of a 3D EBSD tomography experiment in steel. The prior austenite grain microstructure was recovered using reconstruction codes in MTEX. The reconstruction was initially performed on a slice-by-slice basis, and then the microstructure in 3D was constructed via slice to slice linking. Together, this rich data set enables analysis of the 3D microstructure of a high temperature prior austenite domain which contains a twin-boundary, and the relationship of the twin interface to the child variants.

The 3D data presented here indicates that the PAG twin boundary position is not perfectly recovered, but that the recovery of this interface is sufficiently resolved to provide useful and important insight. For example, although some uncertainty remains in its exact location, this does not affect the observed variant relationships or their interpretation. Additionally for the interested reader, further discussion of the twin-boundary position is provided in the Supplementary Material.

The 3D data set enables us to propose mechanisms by which the room temperature microstructure forms, where the twin boundary likely plays an important role in the nucleation and growth of the variants (in addition to any roles that the other prior austenite grain boundaries may play [9], [37] , as well as other defects that can nucleate new variants [38], [39]. To assist in the analysis, we propose the mechanism which is shown schematically in Figure 8. For this analysis and suggested mechanism, we differentiate 'in plane growth' as an analysis of growth which occurs at or near the twin boundary and the growth direction would be along the twin-boundary plane; and 'out of plane growth' as regions where the microstructure extends away from the twin-interface. This builds on prior work in the literature, for example where Hu et al. [40] used laser confocal scanning microscopy to observe the growth of bainite platelets from austenite grain boundaries, as well as sympathetic nucleation of bainite plates (i.e. where two bainite plates grow together, likely due to strain accommodation mechanisms). Note that Hu et al. also observed growth of bainite within grains, but these observations were based only on the surface microstructure, and these plates could either have grown from inside the grain, or from a subsurface structure which was not observed in their analysis. To aid in our destructive and yet 3D analysis, it is useful to note that Hu et al. observe that during cooling, the growth rate of bainite growing at austenite grain boundaries is much higher than the bainite growing from existing bainite (and they infer this is related to the presence of defects near the grain boundaries).

Comparing our analysis to the literature, the work of Taylor et al. [38] also presents surface observations of *in situ* transformations, where so-called bainite "sheaves" (aka laths) form from the high temperature austenite phase which includes both boundary, twin and 'interior' nucleation and growth of the child phase. The surface observations within Taylor et al. indicate that sheaves may nucleate from the grain interior, but these prior literature observations do not preclude the observations and analysis within this manuscript, as Taylor et al. also discuss limitations of their *in situ* measurements which were performed at the sample surface.

Together, this prior literature and the mechanistic uncertainties motivate our 3D analysis, as a 2D section can provide useful information as to the underlying mechanisms which happen within a more complete 3D microstructure, but the 2D data set lacks the 3D connectivity of each variant to the impinging material contained (and constrained) within the entire 3D microstructure. Furthermore, there are related effects which may influence 2D *in situ* studies, such as the presence of the vacuum and free surface, as well as the sectioning effect associated with 2D planar measurements.

For the analysis in the present manuscript, we consider the fully embedded twin-grain set as an entire unit and subsequently split this into sections which correspond to each half of the grain pair, at the twin-plane. This analysis approach is motivated by the fact that we are interested in exploring whether there is likely a relationship between the twin-boundary plane and the observation of microstructural features in the material. To resolve this further, in the schematic in Figure 4 (A) note that the twin plane is split into a grain A and B pair, and Figure 8 (A) shows the nucleation sites within the twin plane, where it is common to see twin related shared position and relative orientations.  Two growth methods for the variants are then shown: Figure 8 (B) highlights in plane growth, which may self terminate, stop at the edge or stop at another variant; and Figure 8 (C) covers out of plane growth, where this growth may self terminate, terminate at the edge of the PAG or another variant. These are discussed in more detail in the following sections.

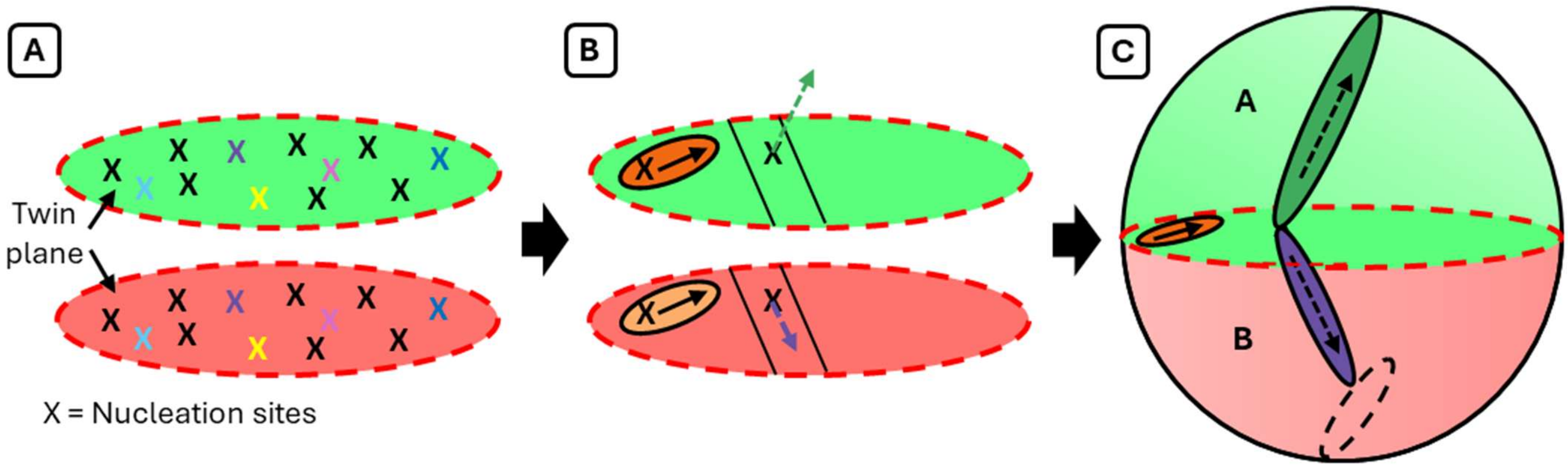

*Figure 8 - Schematic showing how variants nucleate at the twin grain boundary. (A) nucleation sites in the twin planes – common colours indicate twin related nucleation (position & relative orientations); (B) In-plane nucleation and growth; (C) out-of-plane nucleation and growth.*

### 4.1 In-plane growth

In the twin boundary thin layer, certain variants tend to grow within the plane. These variants are typically from the shared packet group and form a flat layer parallel to the twin boundary e.g. with the twin plane in the XY plane, growth is mainly restricted to this XY plane (see also Figure 6, looking at V19-24). Individual variants may self-terminate or stop at another variant (e.g. one that grows out of plane). As seen in Figure 5 (A), the shared packet accounts for ~50% of the voxels in the thin twin boundary region.

Figure 9 summarises data for shared packet variants at the twin boundary, including: an example of the variant growth (Figure 9 (A)), the relative proportions of the variants within the thin boundary region minimum misorientations between the closest variants within the shared packet (Figure 9 (C) and the orientations (Figure 9 (D)).

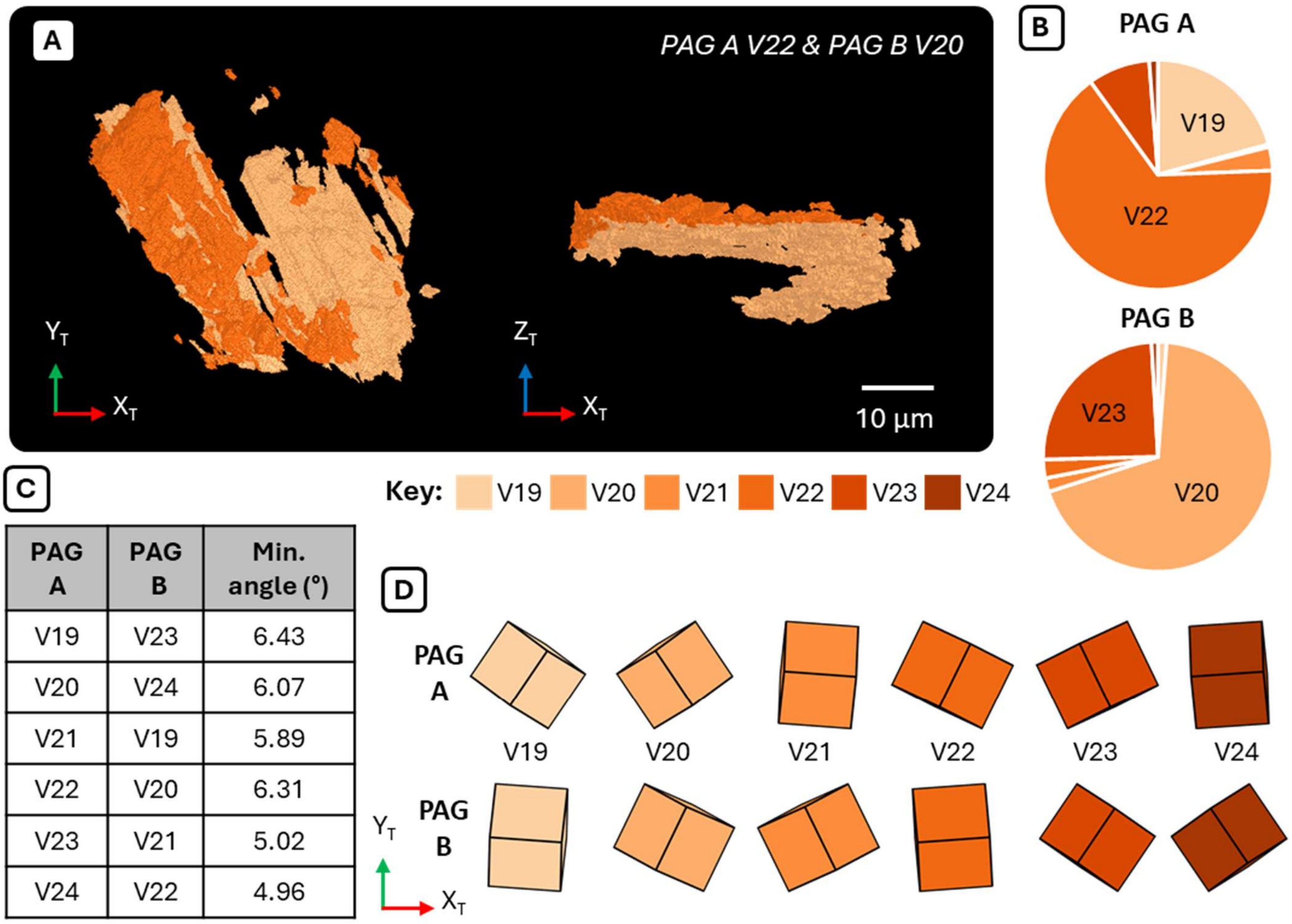

| PAG A | PAG B | Min. angle (°) |
|---|---|---|
| V19 | V23 | 6.43 |
| V20 | V24 | 6.07 |
| V21 | V19 | 5.89 |
| V22 | V20 | 6.31 |
| V23 | V21 | 5.02 |
| V24 | V22 | 4.96 |

*Figure 9 - Shared packet variants – in plane growth. (A) XY and XZ plane views for the largest variant in each PAG; (B) Distribution of the variants in the twin grain boundary region (2µm box) for each PAG; (C) Minimum misorientation angle between variants from each PAG and closest variant ID; (D) orientation cubes for the 6 variants in the shared packet group for each PAG.*

Figure 9 (A) shows the largest shared packet variant (by volume) from each PAG, in both the XY and XZ planes. This data has not been restricted to the thin region at the twin boundary and shows the extent of the growth. These variants (PAG A V22 & PAG B V20) have a misorientation angle of 6.31°, which does give some potential for misidentification of voxels, but this does not seem to have a significant impact here, as the boundary between them is consistent with the twin boundary. The variant (V22) in PAG A is spatially co-located with parts of the variant (V20) in PAG B (which has a higher volume) and both variants appear to form wide, (mostly) flat laths across the twin region, the long axis of which corresponds to the $\langle 110 \rangle_{\gamma}$ in the PAGs / $\langle 111 \rangle_{\alpha}$in the variants. There are also gaps between the variants, where non-shared packet grains nucleate and grow.

The next largest variants for each PAG are also a pair, PAG A V19 & PAG B V23 (not shown), with a similar misorientation angle of 6.43°. In this case, the variants are restricted to a much thinner layer, with no out of plane growth. The variants are again spatially co-located, but this time mostly found at the edges of the layer, with some in plane growth at 60° to the pair seen in (Figure 9 (A)).

Overall, the same trend is observed: sympathetic variant orientations, defined here as the variants in the shared packet in the other grain that have the minimum misorientation (see Figure 9 (C)), are observed for the in-plane growth on either side of the twin boundary. These variants are aligned with $\langle 110 \rangle_{\gamma}$ / $\langle 111 \rangle_{\alpha}$ , and apparently either self terminated or terminated at other variants.

It is also notable that in 3D we observe this in plane growth in the form of a layer at the twin boundary. In a 2D analysis, the presence of the layer and the role of the twin boundary is extremely difficult to understand as the observation is necessarily incomplete (see Figure 4(B)). If the 2D sectioning plane is ~parallel to the twin plane then there is insufficient information to understand the nature of the two twinned parent austenite grains. If the 2D section is ~perpendicular to the twin plane, then there is a risk that the impinging variants obscure the nature of the twin-boundary 'film' like in-plane growth features.

### 4.2 Out of plane growth

Moving to look at out of plane growth, we also see a relationship between the largest variants present in each PAG. This is best demonstrated by looking at one variant pair, as shown in Figure 10. Here, the largest isolated variant region (after connected component analysis in Dragonfly) of variant from the largest variant (by volume) for each PAG is shown. Figure 10 (A) shows the in plane view, with the region cut to just the thin section to look at how they relate at the twin grain boundary. Once again, we see spatial co-location and alignment (approx. $\langle 110 \rangle_{\gamma}$ / $\langle 111 \rangle_{\alpha}$), with sympathetically related orientations for the variants.

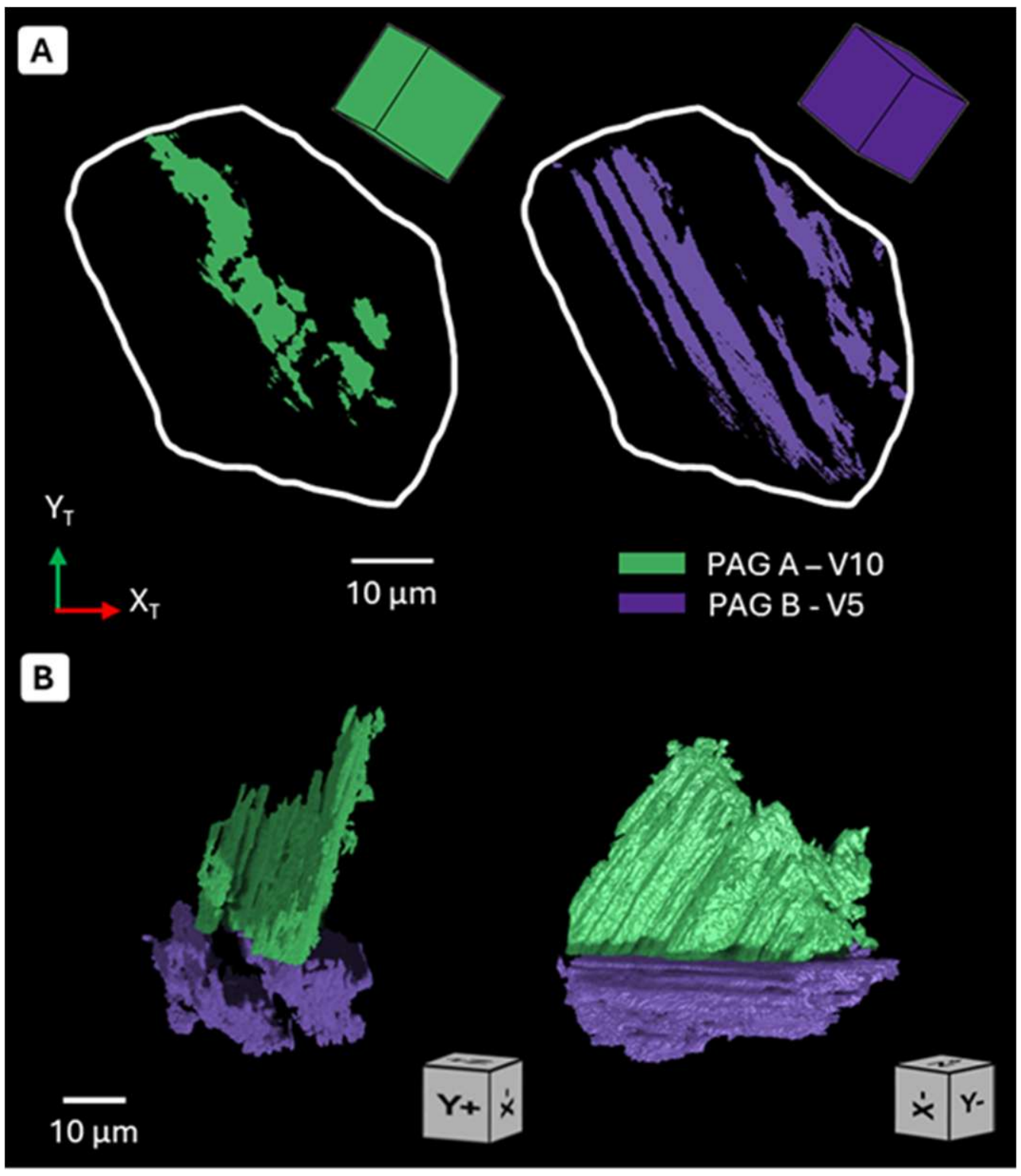

*Figure 10 – Out of plane growth example. Largest isolated group of largest variant (by volume) for each PAG. (A) XY plane above and below the twin plane – a sketched outline is used for the twin plane outline; (B) 3D views*

In Figure 10 (B), the variant pair is shown from two different angles, showing the different shapes of these variants in 3D. The reader is directed to the supplementary material for a video of this pair rotating through 360°, where the need to evaluate variants in 3D becomes far more apparent. The variant in PAG A resembles a thin triangular plane, whilst the variant in PAG A splits into 2 sections, neither of which is obvious from looking at a single direction (or 2D cut). What can also be seen is the alignments of these variants with certain directions, which upon investigation align well with key directions in the PAGs.

The alignment of the variants with crystallographic directions was seen visually in Figure 6 where the 24 variant types were isolated and shown in 3D. This observation was quantified in Figure 7 by analysing these directions, using the principal axes for the isolated variant regions. We find that the long axis directions of the isolated variant regions tend to align with the $\langle 110 \rangle$ in the PAGs, whilst the shortest directions are typically aligned with the $\langle 111 \rangle$ in the PAGs.

### 4.3 Twin boundaries and variant selection

Overall, this analysis reveals that the twin boundary has a strong impact on variant nucleation and growth, and variant selection. This is consistent with earlier 2D observations from low (0.04 wt%) and medium carbon (0.4 wt%) steels which suggested that the formation of annealing twins impacts bainite morphology by (i) blocking the growth of bainitic laths, (ii) providing potential nucleation sites for new laths [12], [13] and (iii) impacting variant selection at the twins [3]. In the present work, by looking at a complete grain in 3D the influence of the PAG twins structure is clearer than a more limited 2D section-based analysis. For variants which have a high aspect ratio and specific shape with respect to each other, the 2D section-based analysis can render an individual variation contribution as a minor feature (e.g. reducing a rod-like variant to a small circular grain). The 3D analysis avoids this, as the volume can be analysed in full and a variety of 2D projections and 3D visualizations performed, in addition to the complete (volume-based) statistical analysis to explore location specific nucleation and growth. It should be noted that there were multiple twinned grains within the volume, consistent with preliminary EBSD studies of the material, and other 2D studies on similar low-medium carbon steels [10], [12]–[14].

The 3D data set here indicates that austenite twins can lead to variant selection and this is probably due to the fact that variant-variant combinations near a twin interface result in specific low & high angle grain boundaries between them (see [36]). For the final (i.e. room temperature) microstructure the nature of the grain orientations, and the grain boundaries, can strongly influence the properties of the final product. Well aligned slip systems, and specific populations of grain boundaries, can impact yield strength, work hardening, and the propagation of cleavage cracks. Further, impact toughness can also be affected [41], as there is evidence that the population of high angle grain boundaries and certain variant pairs have been shown to modify toughness as well as strength [2].

With knowledge of the impact of the twin boundary structure on variant selection, this now provides further opportunities to tailor the final microstructure by changing the number of annealing twins, for example, via small composition adjustments [42], or adjusting the thermomechanical processing route to increase the austenite grain size. Modification of the grain boundary population in FCC-materials is reasonably well explored, and the population of twins (Σ3 boundaries) may be increased using grain boundary engineering [43]. Methods described in the literature include (i) increasing the grain size via extended grain growth where the density of annealing twins increases with grain size [42] leading to more Σ3 boundaries [43], [44] , (ii) the use of iterative deformation and recrystallization steps [45] (for example, tailoring of the multi-step hot rolling process) or (iii) changing the stacking fault energy by alloying, e.g. in some steels, Ni increases and Mn decreases the stacking fault energy [46] where a decrease in stacking fault energy increases the probability of twin formation.

Subdivision of the parent grains with twins increases the number of interfaces and this impacts the variant populations. Smaller regions may see less variants form leading to a coarser microstructure, as we see in PAG B, and seen in literature [3]. For larger twinned regions, where more variants form, this may result in higher local misorientations [3]. Introducing the twin boundaries, we add related variant orientations across the twin boundary, potentially impacting crack paths across the grains.

Ultimately in terms of the design of thermomechanical processing routes for steels, an exploration of the role of the twin-related variant structure could be valuable to influence critical materials properties.

## 5 Conclusions

We have carried out a tomography experiment on a line pipe steel to collect a 3D volume of EBSD data, 150 x 150 x 100 $\mu m^3$ with a 200 $nm^3$ voxel size. The high temperature austenite microstructure was reconstructed from the final microstructure, and this was used to isolate a complete, twinned, PAG within this volume. The twin grain boundary between the grains, and variants present in both grains was then evaluated in more detail, with the following conclusions:

(1) The twin boundary has a strong influence on the variants present within the PAG.
(2) Variants at the twin boundary are often spatially co-located on either side of the boundary and the larger variant pairs have similar orientations.
(3) Some variants grow in-plane at the twin boundary – these are mostly the shared packet variants, and they form a film like layer at the boundary.
(4) Out of plane growth is often related across the twin boundary, including cases with similarly oriented variants with mirrored principal growth directions, and variants may grow to the outer edge of the grain or be stopped by variants growing in from the grain boundary.
(5) The long axis directions of isolated variant regions tend to align with the ⟨110⟩ in the PAG, whilst the shortest directions are typically aligned with the ⟨111⟩ in the PAG.

The value of this analysis lies in reconstructing a full 3D twinned parent austenite grain and its child microstructure, with clearly identified features such as the prior austenite grain boundary and twin plane. As noted by others, statistical analysis of such complex microstructures is challenging; the 3D dataset both enhances and complicates this effort. The quantitative 3D analysis complements the 2D image-based representations used to present our findings and could be extended to larger datasets in future work. Additionally, systematic 2D statistical analysis across multiple sections of a well-defined steel (with controlled chemistry and thermomechanical history) would further broaden the applicability of this work toward a more comprehensive understanding of phase transformations in low-carbon steels.

Analysis of variants within this grain demonstrates that high-temperature twin boundaries have a strong influence on variant selection and growth in the child microstructure. This suggests opportunities to control the final microstructure, and therefore properties, by controlling the twin population in the high-temperature austenite microstructure. Future work will assess the linkage between variant structure and the strength/toughness of high performance line pipe steels.

## 6 Author contributions

**Ruth Birch:** Writing – review & editing, Writing – original draft, Methodology, Investigation, Visualization. **Ben Britton:** Writing – review & editing, Writing – original draft, Supervision, Funding acquisition. **Warren J. Poole:** Writing – review & editing, Supervision, Funding acquisition.

## 7 Data Statement

The 3D dataset is available as raw h5oina files on Zenodo (DOI:10.5281/zenodo.18706015) and high quality figures and videos are available on figshare (DOI:10.6084/m9.figshare.31374256).

## 8 Funding

Electron microscopy was performed at the Electron Microscopy Laboratory (Materials Engineering, UBC), using the AMBER-X pFIB-SEM which was funded by the BCKDF and CFI-IF (39798: AM+). We acknowledge funding from an NSERC Alliance Grant (ALLRP 566973) with Interpro Pipe and Steel and CFI Infrastructure Operating Fund (38798) to support RB. Further, this work was undertaken in part, thanks to the funding from the Canada Research Chair program (WJP and TBB).

## 9 Acknowledgements

The authors would also like to express appreciation and to acknowledge the technical from Interpro Pipe and Steel (formerly EVRAZ NA), in particular Dr. Michael Gaudet. The authors would like to thank Sabyasachi Roy for providing the sample and Professor Matthias Militzer for comments on the research.

This research was enabled in part by support provided by the Digital Research Alliance of Canada, formerly Compute Canada, and the Cedar cluster (now retired) located at Simon Fraser University.

# 1 Supplementary Material

## 1.1 Sample Details

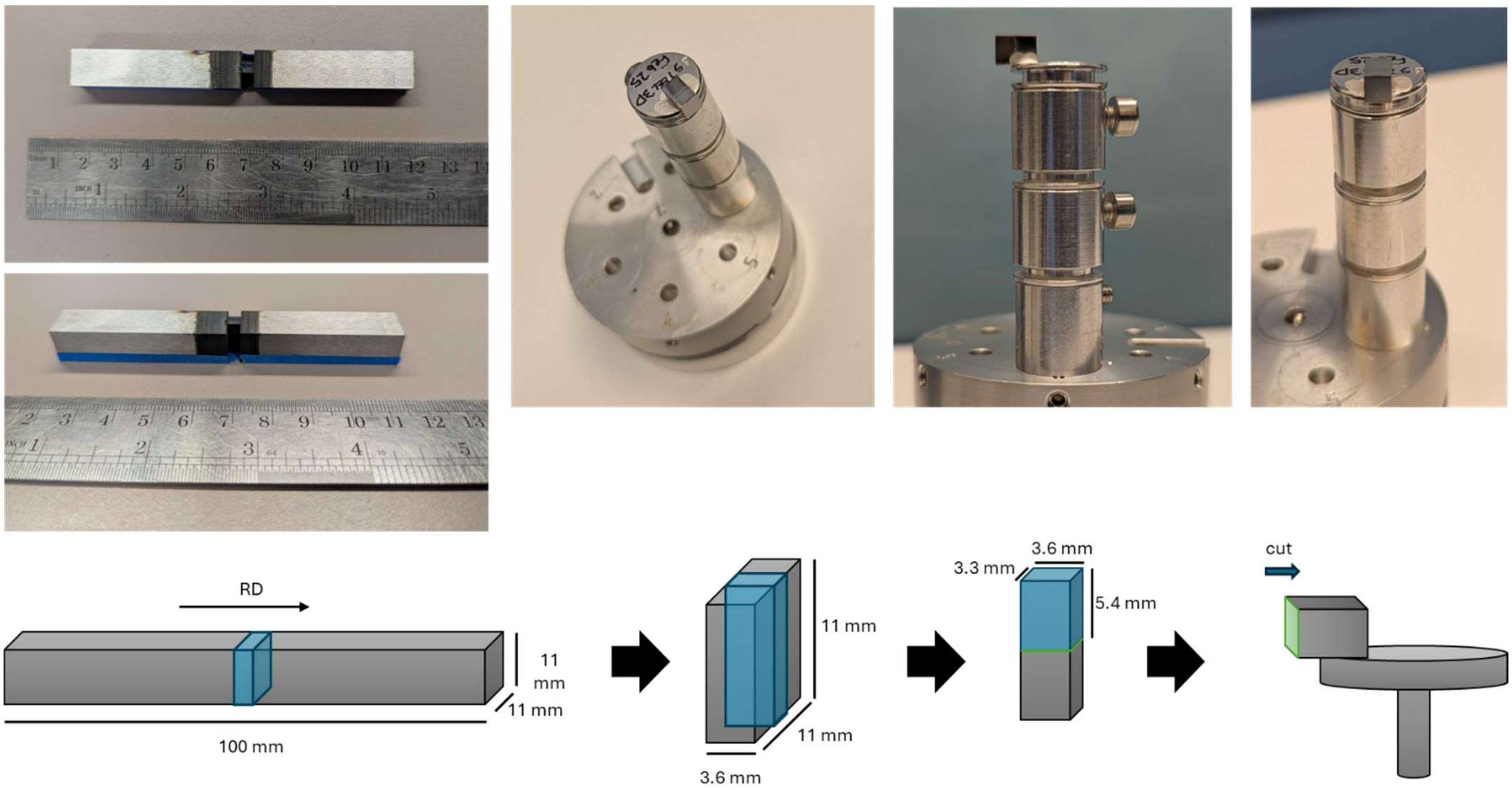

*Figure S1: Sample details - Tomography cut relative to the sample for the heat treatment.*

## 1.2 Surface EBSD

Preliminary EBSD of a representative area of the sample surface prior to the tomography experiment is shown in Figure S2. This map was collected using a voltage of 20kV and current of 10 nA for an area of 1 mm x 1 mm, at a step size 0.2 μm. Patterns were captured at 156 x 128 pixels with an exposure time of 2.5 ms, and processed online in Oxford Instruments AZtec software, with a minimum of 6 bands required to index the point.

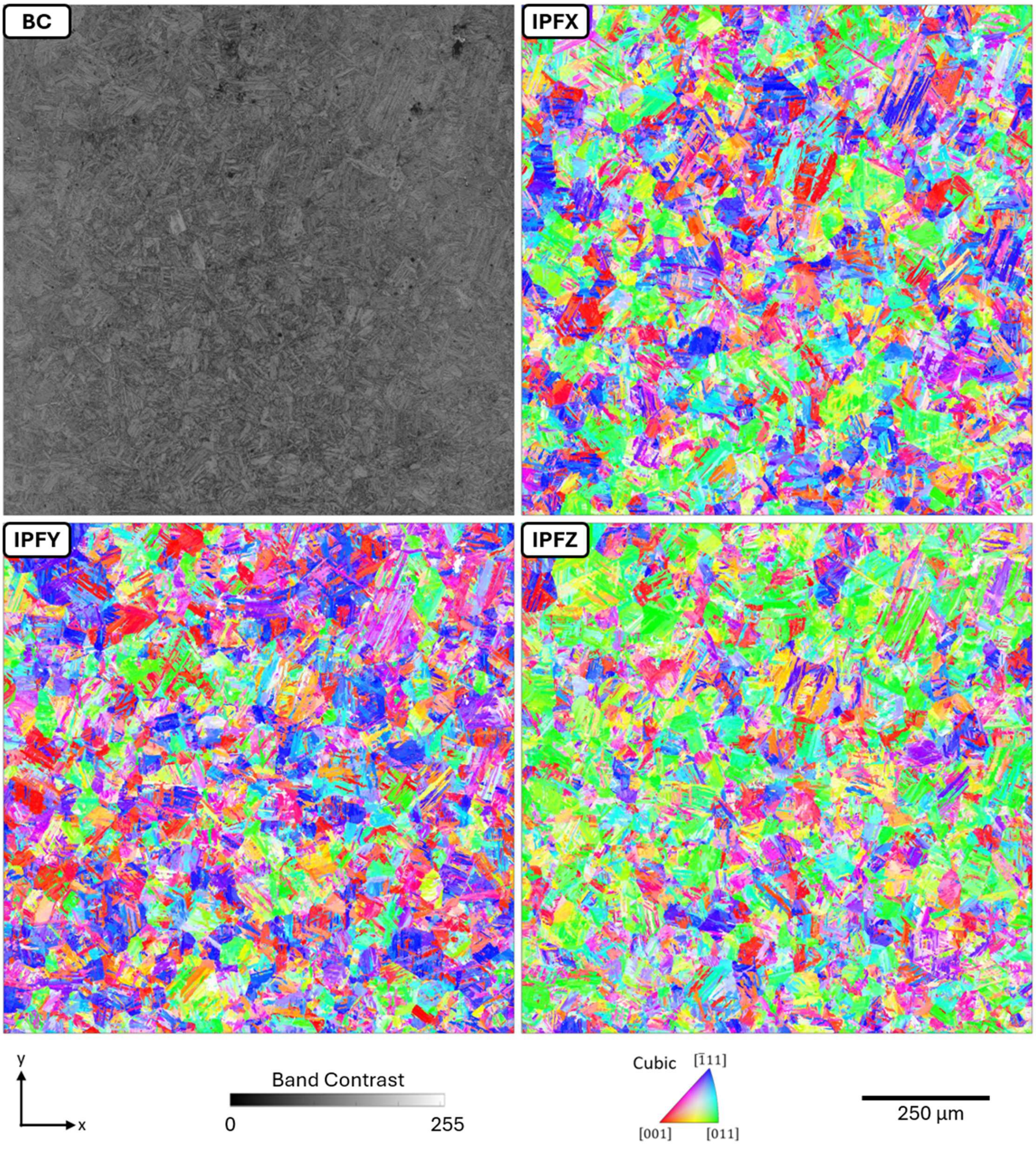

*Figure S2: 1mm$^2$ EBSD mapping of top surface, including band contrast (BC) and inverse pole figure maps in X,Y and Z.*

### 1.3 Twin Position

For context, the reader is reminded that misassignment of the variants to the ‘wrong’ parent twin orientation is primarily important for the ‘shared’ variant pairs which are inherited through the twin-OR and the KS-OR (e.g. variant 19 from Twin A and variant 23 from Twin B). In these cases, the location of the twin interface can be located in or at the edge of this variant grouping and misassignment could result in a loss in precision of the exact twin position.

The problem of misassignment occurs as it is challenging to differentiate between variants that are shared by both parent grains across a twin boundary. In fact, if it is an exact twin pair, adhering strictly to the orientation relationship (OR) e.g. Kurdjumov-Sachs (KS), then it is not possible to determine to which PAG the points/child grains belong based on the EBSD orientations. You could take into account other factors, such as backscatter (or forescatter) imaging data, or the band contrast, all of which may give an indicator of where the boundaries should be placed, but this would be computationally inefficient for large volumes and potentially introduce other errors.

From our steel analysis and using the iterative code developed by Nyyssönen [1], we often observe that the best fitting OR to enable prior austenite grain reconstruction deviates from the ideal KS OR. The best fitting OR is determined using an iterative, best-fitting approach to the experimental data. When this iterative OR is used, there is a misorientation between the shared variant pairs and this allows the child grains/voxels to be assigned to one the twinned PAGs (least misorientation angle to one of the calculated child variants = PAG) with greater confidence.

In our case, we used a mean iterative OR for the volume, which deviates from the KS OR as shown in Figure S3. Furthermore, analysis of the recovered parent austenite grain twin pair misorientation also indicates a ~0.5° from a strict 60° <111> twin relationship (this is consistent with experimental work by [2]). Together, these result in a deviation of the (child) variant-variant misorientations, with the smallest variant-variant misorientations being ~5°. The variant-variant misorientations using the mean iterative OR for this PAG pair are detailed in Supplementary Table S1.

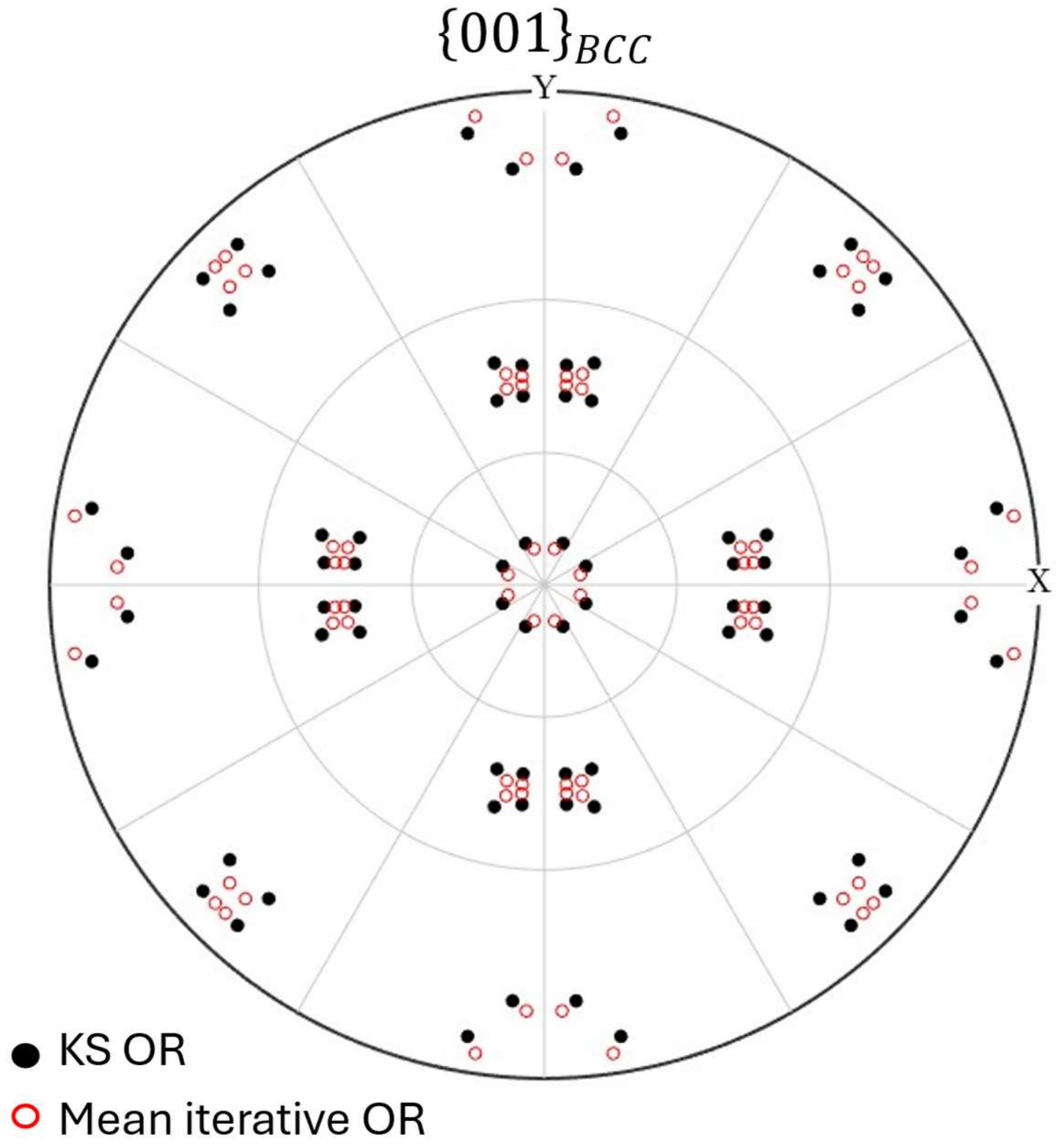

*Figure S3 - Comparison of Kurdjumov Sachs (KS) orientation relationship (OR) and mean iterative OR used for the dataset. Pole figure for calculated variants from an $(0,0,0)_{FCC}$ parent orientation for each case.*

The 3D reconstruction is still a challenge, due to experimental error and uncertainty. For example, limitations in accuracy in EBSD data collection (this is indexed using the 'optimized BD' Hough transform method in Aztec, which has a precision of between 0.5° and 0.1°) and thresholds used within the reconstruction code (e.g. for grain identification or for triggering reidentification of the PAG orientation due to uncertain orientations present (i.e. high uncertainty) to see if they fit better, and within a set threshold with a surrounding PAG – this step helps resolve the presence of any island grains that may result due to the common child variants between twins – see [3]).

Together these effects mean that it can be valuable to scrutinize the location of the final boundary position in these complicated microstructures, especially when the analysis focusses on the twin habit plane.

To provide context on the expected microstructure in these alloys, a coherent Σ3 boundary is typically planar/flat with a misorientation of 60° about <111> and the interfacial plane is the {111} plane perpendicular to the rotation axis. An incoherent Σ3 boundary also has a misorientation of 60° about <111>, but contains steps or ledges (and other defects) and this results in a non-planar interface. Deviations away from the ideal misorientation can occur due to deformation or lattice rotation that happens after the twin has formed. In this steel, the Σ3 are likely annealing twins.

In practice, the final the position of the twin boundary as presented has a mix of easy to identify 'flat' boundary; as well as sections where the exact position is more uncertain, likely due to difficulties with the shared variants from the twin. For the parent grain twin boundary, and the subsequent variant analysis, the recovered twin interface can also be affected by variant nucleation and growth, where variants from one side of the twin impinge and cross over across the original twin grain boundary.

In the data collected via EBSD, the exact location of the twin interface can be obtained from the reconstructed parent data (as described prior), but also it can be contained within other crystallographic signatures such as retained austenite (as identified in [3]) or the band contrast data. Retained austenite is often not available in FIB-based studies due to FIB-induced phase transformations [4] or related artifacts.

For the 3D data set in the present work, the twin interface was recovered with sufficient confidence for the analyses presented, as demonstrated by the following case studies from selected slices through the PAG twin grain boundary shown in S3.

### 1.3.1 Case 1 – straight boundary (upper set of images in Figure S4)

These are the sections that behave as we would expect from literature, with a relatively straight twin boundary. The child orientations (and therefore variants) on either side of the twin boundary are sufficiently different to separate between the two potential PAGs with the code. The initial microstructure and band contrast support the position positioning of the PAG twin boundary i.e. there are child grain boundaries in the same place, and it matches with the darker area (lower quality) on the band contrast that may be associated with a grain boundary.

### 1.3.2 Case 2 – uncertain boundary sections (lower set of images in Figure S4)

There is uncertainty of the location of the twin interface as voxels near the interface can be associated with either identified parent austenite grain. The twin interface in the reconstructed data is not flat, and the backscatter also does not show a flat interface. This is likely due to the shared variants from a twin problem. In the case shown in figure S4, the shared packet variants are in both cases are shown in orange (i.e. V19-24) and we see that where the boundary (see the PAG map for twin boundary position) deviates from straight, the variants in this region are indeed from this shared packet group.

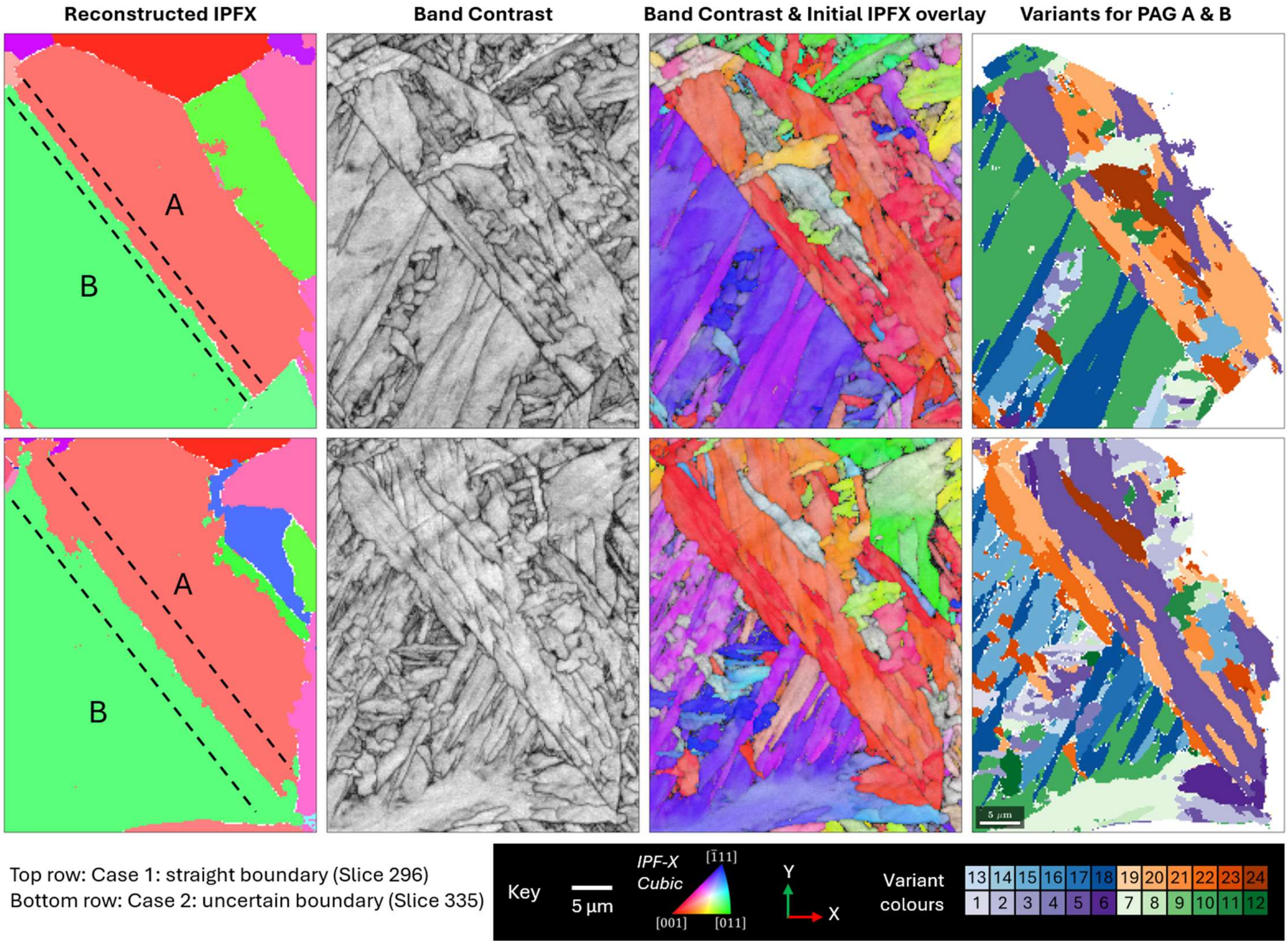

*Figure S4: Twin boundary examples – Two cases addressing the impact of variants with similar orientations on the final position of the PAG twin boundary. Reconstructed (PAG) microstructure – IPFX, band contrast, band contrast with IPFX original overlaid and variants for PAG A & B (24 variant colours) are shown in each case. Top row: case 1 – straight twin boundary; Bottom row – uncertain (wavy) twin boundary. Cropped regions from the original slices are shown.*

Quantitatively, this will provide a small uncertainty on the number of voxels attributed to each parent grain and the specific variant numbers due to misclassification of the parent grain identity. Qualitatively, correlative analysis of field plots like band contrast together with crystallographic orientation and grain identification can be used to provide more unambiguous identification of the twin interface location. However, automated methods to use this correlative data are challenging to implement in a full 3D algorithm at present.

Overall, although unambiguous identification of the twin-boundary position in the parent microstructure would be valuable, the automated approach using the 2.5D re-sort algorithm enables sufficient identification of the twin location for variant analysis relative to the twin interface.

| PAG A / PAG B | V1 | V2 | V3 | V4 | V5 | V6 | V7 | V8 | V9 | V10 | V11 | V12 | V13 | V14 | V15 | V16 | V17 | V18 | V19 | V20 | V21 | V22 | V23 | V24 |
|---|---|---|---|---|---|---|---|---|---|---|---|---|---|---|---|---|---|---|---|---|---|---|---|---|
| V1 | 45.4 | 36.8 | 32.2 | 53.4 | 40.9 | 24.5 | 36.4 | 45.5 | 32.2 | 44.3 | 49.8 | 23.5 | 50.3 | 35.0 | 52.3 | 53.2 | 34.5 | 47.3 | 20.0 | 46.1 | 54.9 | 20.3 | 50.9 | 55.1 |
| V2 | 32.2 | 47.0 | 46.8 | 32.3 | 52.3 | 49.4 | 44.3 | 30.6 | 48.0 | 36.8 | 22.2 | 51.7 | 29.1 | 50.8 | 41.5 | 21.9 | 48.3 | 37.7 | 51.0 | 50.9 | 19.2 | 56.8 | 46.7 | 17.4 |
| V3 | 44.1 | 31.8 | 47.2 | 36.6 | 23.5 | 50.9 | 33.4 | 46.9 | 46.2 | 33.3 | 52.3 | 48.9 | 37.6 | 47.6 | 22.6 | 41.8 | 50.2 | 29.9 | 56.0 | 20.3 | 51.0 | 50.6 | 18.2 | 46.9 |
| V4 | 37.6 | 44.2 | 32.1 | 45.4 | 48.5 | 23.6 | 44.2 | 38.0 | 31.8 | 52.2 | 42.2 | 24.0 | 48.5 | 34.4 | 53.3 | 53.6 | 34.7 | 50.6 | 20.1 | 53.6 | 47.4 | 19.7 | 55.2 | 52.2 |
| V5 | 23.7 | 50.8 | 49.3 | 24.7 | 55.2 | 51.0 | 48.8 | 22.4 | 53.4 | 41.2 | 14.4 | 56.3 | 21.9 | 53.6 | 44.0 | 16.2 | 53.4 | 41.8 | 51.7 | 53.1 | 12.4 | 59.5 | 49.0 | 11.2 |
| V6 | 48.4 | 23.5 | 52.4 | 40.8 | 15.5 | 55.4 | 24.8 | 50.4 | 48.6 | 25.5 | 55.1 | 50.3 | 41.5 | 52.5 | 16.4 | 44.0 | 52.9 | 22.5 | 59.0 | 13.1 | 53.1 | 51.1 | 11.5 | 49.0 |
| V7 | 34.6 | 47.4 | 50.8 | 35.2 | 52.4 | 53.7 | 44.4 | 32.2 | 45.3 | 36.4 | 23.6 | 49.8 | 32.4 | 53.1 | 40.9 | 24.7 | 45.2 | 36.7 | 55.5 | 51.0 | 20.3 | 54.9 | 46.0 | 20.2 |
| V8 | 48.5 | 37.8 | 28.9 | 50.4 | 41.6 | 21.7 | 36.8 | 48.2 | 30.6 | 44.3 | 51.9 | 22.3 | 46.4 | 32.2 | 52.3 | 48.9 | 32.2 | 47.0 | 17.2 | 46.8 | 56.9 | 19.3 | 50.9 | 50.5 |
| V9 | 50.6 | 29.9 | 37.8 | 47.2 | 22.4 | 42.0 | 33.2 | 46.6 | 47.2 | 33.3 | 49.2 | 52.6 | 46.8 | 36.9 | 23.3 | 50.6 | 44.4 | 31.6 | 47.1 | 18.1 | 51.0 | 51.3 | 20.1 | 55.7 |
| V10 | 34.8 | 50.1 | 48.3 | 34.6 | 52.9 | 53.4 | 52.3 | 31.9 | 37.9 | 44.3 | 24.1 | 42.2 | 32.4 | 45.2 | 48.5 | 23.8 | 37.4 | 44.3 | 52.1 | 54.8 | 19.8 | 47.3 | 53.6 | 20.3 |
| V11 | 53.6 | 41.9 | 21.7 | 53.3 | 44.1 | 16.0 | 41.2 | 53.6 | 22.4 | 48.9 | 56.4 | 14.5 | 49.0 | 24.6 | 55.2 | 50.6 | 23.7 | 50.8 | 11.0 | 49.2 | 59.1 | 12.5 | 53.2 | 51.2 |
| V12 | 53.3 | 22.5 | 41.7 | 52.2 | 16.3 | 44.1 | 25.4 | 49.0 | 50.7 | 24.7 | 50.7 | 55.3 | 52.1 | 41.0 | 15.2 | 55.1 | 48.7 | 23.3 | 49.2 | 11.3 | 51.5 | 53.3 | 12.8 | 59.4 |
| V13 | 38.0 | 47.0 | 30.4 | 45.5 | 50.6 | 22.2 | 47.4 | 38.9 | 28.5 | 50.7 | 42.9 | 21.0 | 48.2 | 32.0 | 49.0 | 53.5 | 31.8 | 46.6 | 18.9 | 55.7 | 48.0 | 16.7 | 50.6 | 52.2 |
| V14 | 44.1 | 33.4 | 44.5 | 36.2 | 24.8 | 49.0 | 35.7 | 47.2 | 50.3 | 36.1 | 52.3 | 53.3 | 36.7 | 44.5 | 25.5 | 41.1 | 52.4 | 33.2 | 54.1 | 21.5 | 51.1 | 55.2 | 21.1 | 46.2 |
| V15 | 24.1 | 48.8 | 51.7 | 23.7 | 50.3 | 56.3 | 53.2 | 21.1 | 42.9 | 52.5 | 15.1 | 45.3 | 22.4 | 49.7 | 55.2 | 14.7 | 42.1 | 52.4 | 54.3 | 51.0 | 10.1 | 50.3 | 58.9 | 12.2 |
| V16 | 42.3 | 52.5 | 22.1 | 49.9 | 55.2 | 14.3 | 52.5 | 42.9 | 21.1 | 53.7 | 45.3 | 15.0 | 51.9 | 23.4 | 50.7 | 56.4 | 24.0 | 49.3 | 11.8 | 59.4 | 50.3 | 10.0 | 51.4 | 54.4 |
| V17 | 52.1 | 33.2 | 36.9 | 44.1 | 25.4 | 41.3 | 36.1 | 50.6 | 47.6 | 35.7 | 53.6 | 52.7 | 44.2 | 36.6 | 24.7 | 48.7 | 44.4 | 33.3 | 46.5 | 21.0 | 55.6 | 51.4 | 21.3 | 53.8 |
| V18 | 32.0 | 46.2 | 48.0 | 32.3 | 48.6 | 53.4 | 50.2 | 28.6 | 38.9 | 47.5 | 21.2 | 42.9 | 30.7 | 45.3 | 50.6 | 22.6 | 37.9 | 47.1 | 52.1 | 50.3 | 16.9 | 48.0 | 55.7 | 19.3 |
| V19 | 53.5 | 20.3 | 51.0 | 45.9 | 13.1 | 53.3 | 21.4 | 55.5 | 50.2 | 21.1 | 59.4 | 51.0 | 46.6 | 51.1 | 11.4 | 49.0 | 54.8 | 18.1 | 58.3 | 9.8 | 58.2 | 51.7 | 6.4 | 54.1 |
| V20 | 20.2 | 55.8 | 50.9 | 20.2 | 59.0 | 51.7 | 53.9 | 19.0 | 52.1 | 46.3 | 11.9 | 54.3 | 17.4 | 55.5 | 49.1 | 11.2 | 52.1 | 46.9 | 52.2 | 58.2 | 8.9 | 59.4 | 54.1 | 6.1 |
| V21 | 52.3 | 47.1 | 17.2 | 55.1 | 49.2 | 11.0 | 46.4 | 52.2 | 19.1 | 53.9 | 54.4 | 12.1 | 50.6 | 20.1 | 59.4 | 51.3 | 20.2 | 55.9 | 5.9 | 54.2 | 59.5 | 9.2 | 58.3 | 51.8 |
| V22 | 55.2 | 18.1 | 46.8 | 50.8 | 11.3 | 49.2 | 21.0 | 50.7 | 55.8 | 21.3 | 51.4 | 59.0 | 50.8 | 46.1 | 12.9 | 53.0 | 53.8 | 20.1 | 54.3 | 6.3 | 52.1 | 58.4 | 9.5 | 58.1 |
| V23 | 19.8 | 50.6 | 56.8 | 20.4 | 51.1 | 59.5 | 55.2 | 16.8 | 48.0 | 51.2 | 10.0 | 50.3 | 19.4 | 54.8 | 53.2 | 12.7 | 47.3 | 51.1 | 59.4 | 51.7 | 5.0 | 55.4 | 58.2 | 9.3 |
| V24 | 47.4 | 51.2 | 19.0 | 55.0 | 53.3 | 12.2 | 51.3 | 48.1 | 16.8 | 55.6 | 50.3 | 10.0 | 57.0 | 20.2 | 51.5 | 59.1 | 19.6 | 51.0 | 8.8 | 58.3 | 55.4 | 5.0 | 52.1 | 59.5 |

*Table S1: Misorientation angle (in degrees) between PAG A variants and PAG B variants, calculated using the mean iterative OR for the volume and the mean orientation for each PAG.*

### 1.4 Variant Composition – Packet groups (statistics)

Statisitics for packet groups in thin **twin boundary region** (regions > 50 voxels included)

Total voxels considered in PAG A region: 176047

Total voxels considered in PAG B region: 163087

| PAG | Packet | | Voxels | % of total (2µm box) |
|---|---|---|---|---|
| A | 1 | V1-6 | 27981 | 15.9% |
| | 2 | V7-12 | 34114 | 19.4% |
| | 3 | V13-18 | 34780 | 19.8% |
| | 4 | V19-24 | 79172 | 45.0% |
| B | 1 | V1-6 | 54976 | 33.7% |
| | 2 | V7-12 | 15071 | 9.2% |
| | 3 | V13-18 | 9075 | 5.6% |
| | 4 | V19-24 | 83965 | 51.5% |

Statisitics for packet groups in **complete PAG** (regions > 50 voxels included)

Total voxels considered in PAG A: 4984565

Total voxels considered in PAG B: 1868047

| PAG | Packet | | Voxels | % of total (complete PAG) |
|---|---|---|---|---|
| A | 1 | V1-6 | 1031800 | 20.7% |
| | 2 | V7-12 | 1560587 | 31.3% |
| | 3 | V13-18 | 1732999 | 34.8% |
| | 4 | V19-24 | 659179 | 13.2% |
| B | 1 | V1-6 | 873052 | 46.7% |
| | 2 | V7-12 | 189109 | 10.1% |
| | 3 | V13-18 | 121991 | 6.5% |
| | 4 | V19-24 | 683895 | 36.6% |

### 1.5 Variant composition – Packet groups

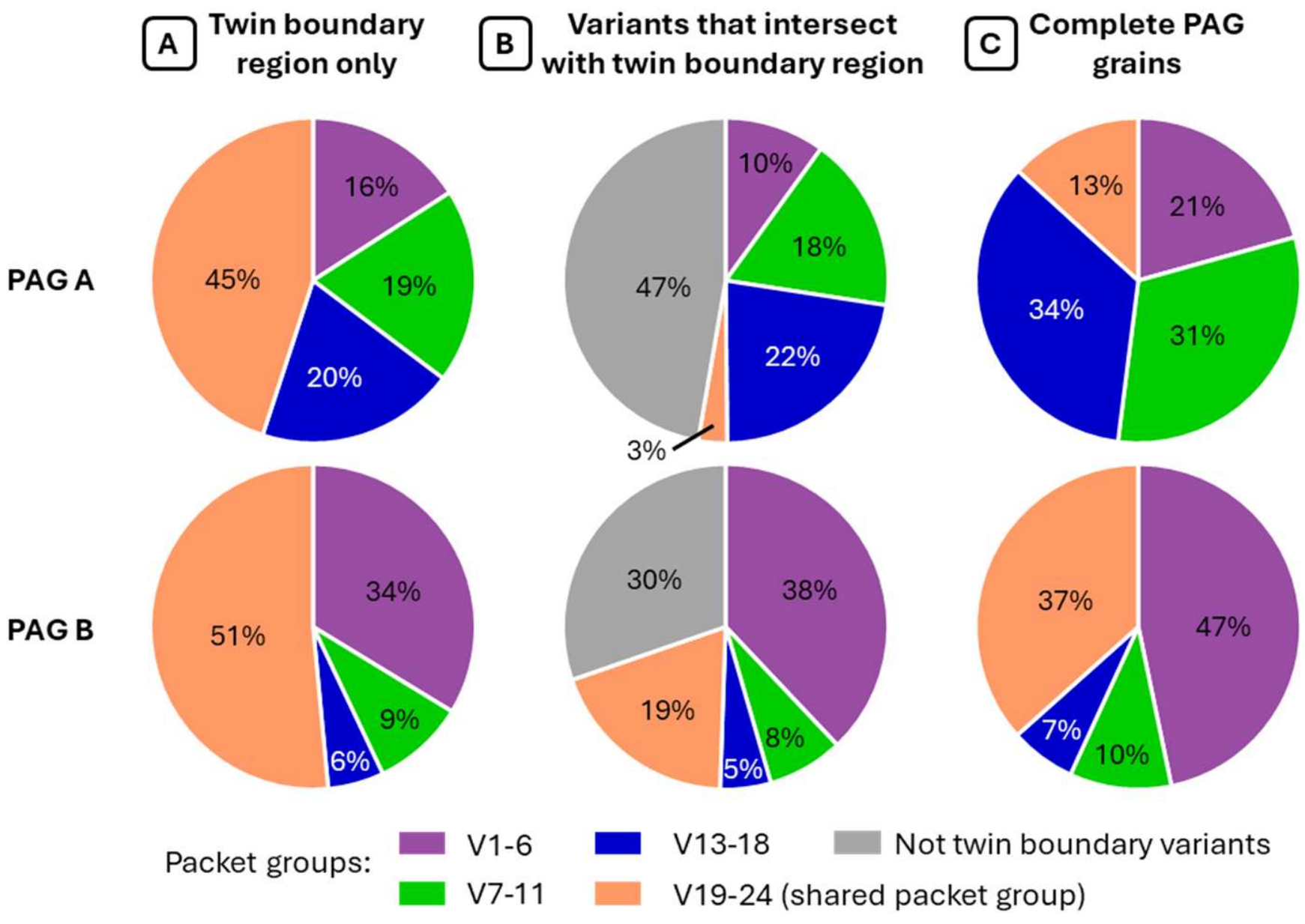

*Figure S5: Variant distribution into packet groups. (A) twin boundary region only – 2 µm box around the twin boundary; (B) variants intersecting with the twin boundary region, including points not associated with the twin boundary ('not twin boundary variants' – grey); and (C) in the complete PAG grain volume. Variants containing a minimum of 50 voxels included. Total voxels considered: PAG A: 5,192,442 (complete grain) / 176,047 (twin boundary region); PAG B: 1,916,858*

*(complete grain) / 163,087 (twin boundary region). Variant numbers are specific to the PAG grain – only the shared packet group variants are common (due to symmetry) and can be read across.*

### 1.6 Variant composition – Bain groups

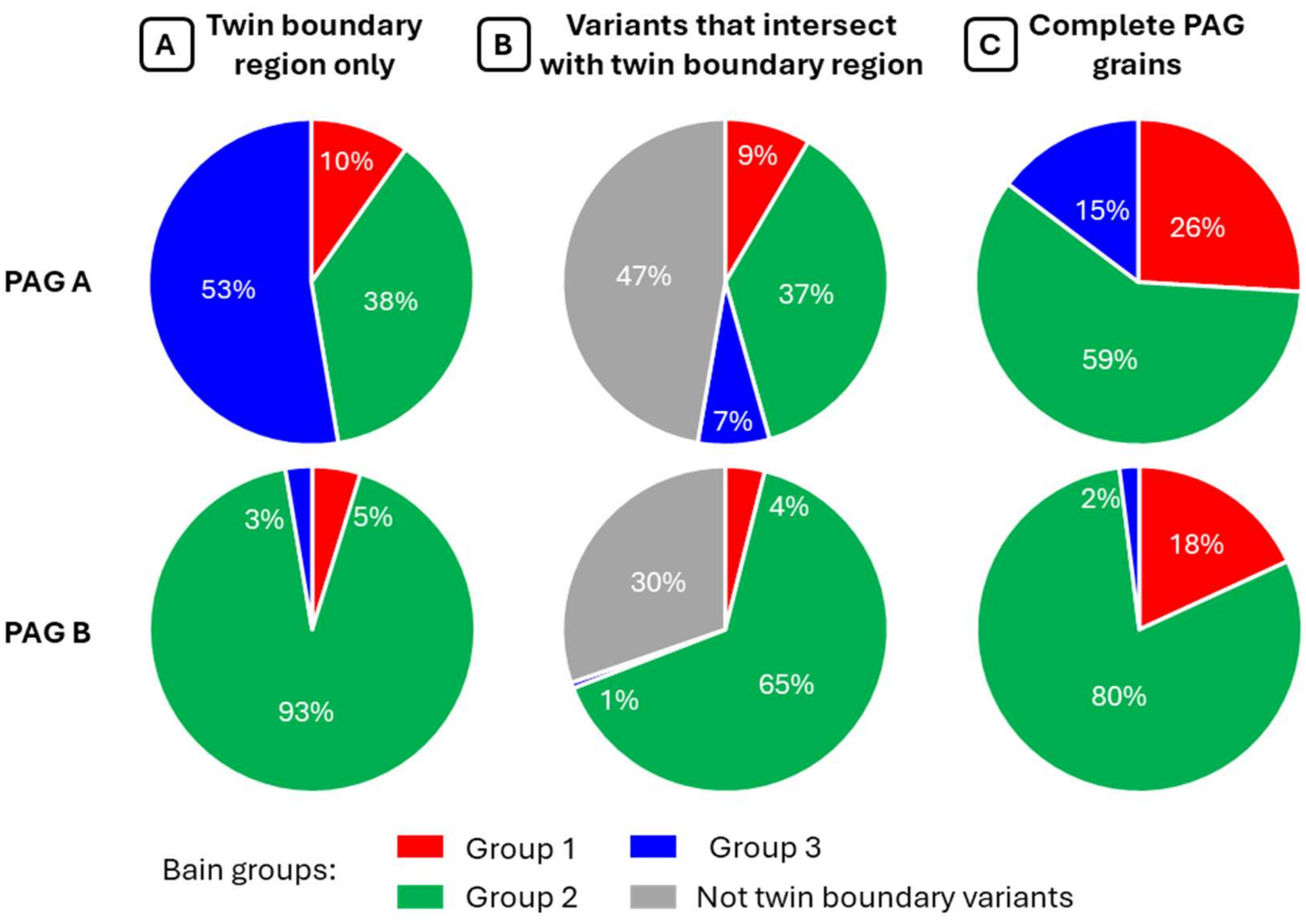

*Figure S6: Variant distribution into Bain groups. (A) twin boundary region only – 2 µm box around the twin boundary; (B) variants intersecting with the twin boundary region, including points not associated with the twin boundary ('not twin boundary variants' – grey); and (C) in the complete PAG grain volume. Variants containing a minimum of 50 voxels included. Total voxels considered: PAG A: 5,192,442 (complete grain) / 176,047 (twin boundary region); PAG B: 1,916,858 (complete grain) / 163,087 (twin boundary region).*

### 1.7 Variant composition – individual variants

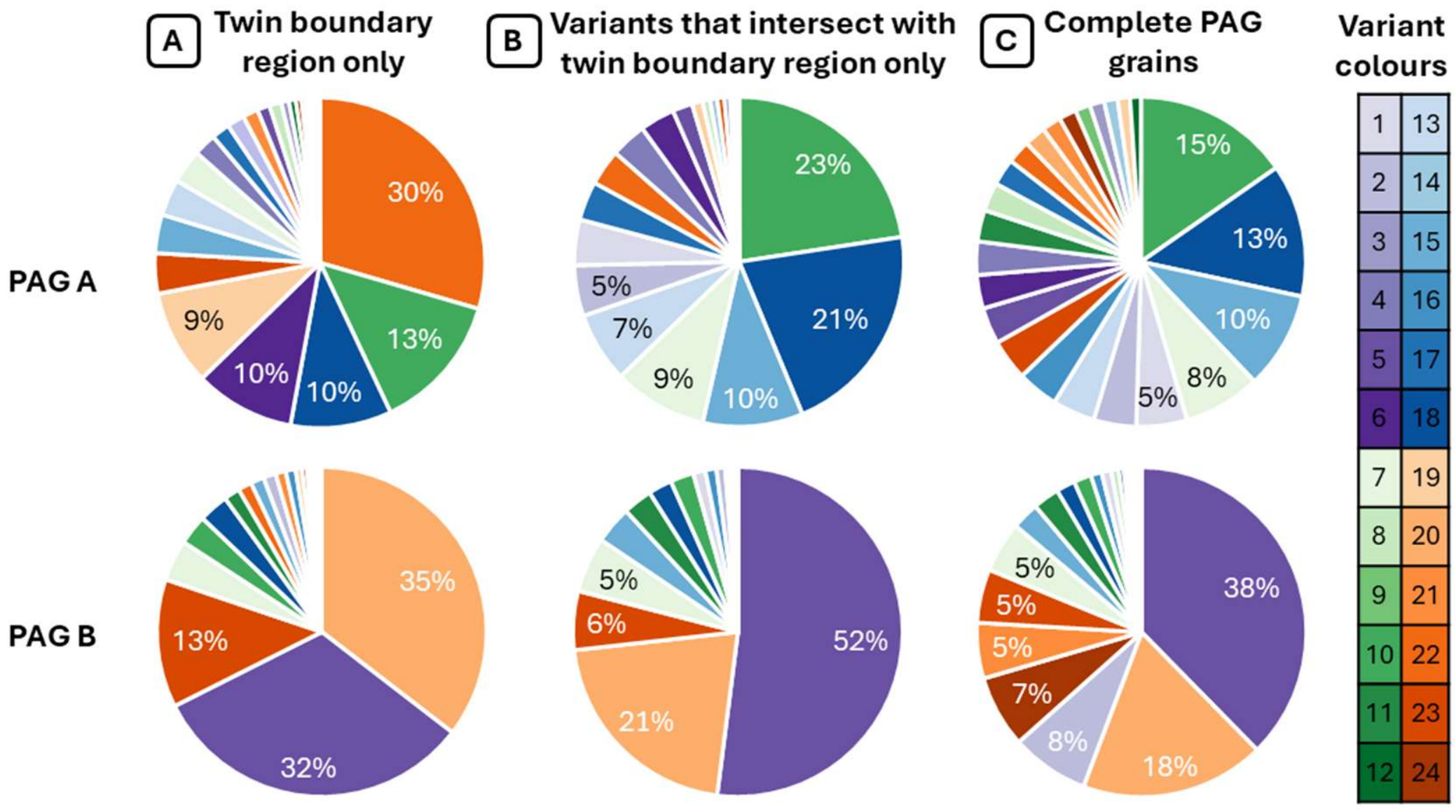

*Figure S7: Variant distribution into individual variant groups. (A) twin boundary region only – 2 µm box around the twin boundary; (B) variants intersecting with the twin boundary region, including points not associated with the twin*

*boundary ('not twin boundary variants' – grey); and (C) in the complete PAG grain volume. Variants containing a minimum of 50 voxels included. Total voxels considered: PAG A: 5,192,442 (complete grain) / 176,047 (twin boundary region); PAG B: 1,916,858 (complete grain) / 163,087 (twin boundary region). Note: all segments >= 5% are labelled*

### 1.8 Videos and high quality images

Videos and high quality images for the figures in the paper are available at: DOI:10.6084/m9.figshare.31374256